\documentclass[aps,prl,superscriptaddress,amsmath,amssymb,floatfix,preprint]{revtex4}
\usepackage{graphicx}
\usepackage{times}
\usepackage{amsmath}
\usepackage{amssymb}
\usepackage{float}
\def\d{\delta}

\def\b{\beta}

\def\m{\mu}
\def\t{\tau}

\begin{document}
\vspace*{0.2in}

\begin{flushleft}
{\Large
\textbf\newline{Optimal number of spacers in CRISPR arrays} % Please use "sentence case" for title and headings (capitalize only the first word in a title (or heading), the first word in a subtitle (or subheading), and any proper nouns).
}
\newline
\\
Alexander Martynov\textsuperscript{1*},
Konstantin Severinov\textsuperscript{1,2,3},
Yaroslav Ispolatov\textsuperscript{4*},
\\

\bigskip
\textbf{1} Center for Data-Intensive Biomedicine and Biotechnology, Skolkovo Institute of Science and Technology, Moscow, Russia
\\
\textbf{2} Waksman Institute of Microbiology, Rutgers, The State University of New Jersey, Piscataway, NJ, USA
\\
\textbf{3} Institute of Molecular Genetics, Russian Academy of Sciences, Moscow, Russia
\\
\textbf{4} Department of Physics, University of Santiago de Chile, Santiago, Chile
\\
\bigskip

% Insert additional author notes using the symbols described below. Insert symbol callouts after author names as necessary.

% Use the asterisk to denote corresponding authorship and provide email address in note below.
* YI jaros007@gmail.com, AM alexander.martynov@skolkovotech.ru

\end{flushleft}
\section*{Abstract}
We estimate the number of spacers in a CRISPR array of a bacterium which maximizes its protection against a viral attack. The optimality follows from a competition between two trends: too few distinct spacers make the bacteria vulnerable to an attack by a virus with mutated corresponding protospacers, while an excessive variety of spacers dilutes the number of the CRISPR complexes armed with the most recent and thus most effective spacers. We first evaluate the optimal number of spacers in a simple scenario of an infection by a single viral species and later consider a more general case of multiple viral species. We find that depending on such parameters as the concentration of CRISPR-CAS interference complexes and its preference to arm with more recently acquired spacers, the rate of viral mutation,  and the number of viral species, the predicted optimal array length lies within a range quite reasonable from the viewpoint of recent experiments.

% Please keep the Author Summary between 150 and 200 words
% Use first person. PLOS ONE authors please skip this step. 
% Author Summary not valid for PLOS ONE submissions.   
\section*{Author summary}
CRISPR-Cas system is an adaptive immunity defense in bacteria and archaea against viruses. It works by accumulating in bacterial genome an array of spacers, or fragments of virus DNA from previous attacks. By matching spacers to corresponding parts of virus DNA called protospacers, CRISPR-Cas system identifies and destroys intruder DNA. Here we theoretically estimate the number of spacers that maximizes bacterial survival. This optimum emerges from a competition between two trends: More spacers allow a bacterium to hedge against mutations in viral protospacers. However, keeping too many spacers makes the older ones inefficient because of  accumulation of mutations in corresponding protospacers in viruses. Thus, fewer CRISPR-Cas molecular machines are left armed with more efficient young spacers. We have shown that a higher efficiency of CRISPR-Cas system allows a bacterium to utilize more spacers, increasing the optimal array length. On contrary, a higher viral mutation rate makes older spacers useless and favors shorter arrays. A higher diversity in viral species reduces the efficiency of CRISPR-Cas  but does not necessary lead to longer arrays. We think that our study provides a new viewpoint at a huge variety in the observed array lengths and adds relevance to evolutionary models of bacterial-phage coexistence.

% Use "Eq" instead of "Equation" for equation citations.
\section{Introduction}
CRISPR-Cas systems provide prokaryotes with adaptive immunity against viruses and plasmids by targeting foreign nucleic acids \cite{Makarova2006, Bolotin2005, Barrangou2007}. Multiple CRISPR-Cas systems differing in molecular mechanisms of foreign nucleic acids destruction, cas genes, CRISPR repeats structure, and the lengths, numbers and origin of spacers have been discovered \cite{Makarova2012,Makarova2015}. Yet the current understanding of diversity and function of CRISPR-Cas systems is far from being complete. The origins and, therefore, the targets of most spacers remain unknown \cite{Hargreaves2014, McGhee2012, VanBelkum2015}. The ubiquity of CRISPR-Cas systems in archaea compared to less than 50\% presence in bacteria is also not well-explained \cite{Makarova2012,Makarova2013}. Evolutionary reasons for plethora of distinct CRISPR-Cas systems types, often coexisting in the same genome, remain largely unexplored \cite{Makarova2015,Agari2010,Rath2015}. It is also not clear why CRISPR arrays of some CRISPR-Cas systems contain only one or few spacers, while others have dozens or even hundreds of them \cite{Diez-Villasenor2010, Horvath2008, Grissa2007, Hale2008, Agari2010, Rath2015}.  It is commonly accepted that the number of spacers in an array is a result of a compromise between better protection offered against abundant, diverse, and faster evolving viruses by a larger spacer repertoire and a higher physiological cost of maintaining a longer array \cite{Levin2013}. However, even the largest of the CRISPR systems contribute only 1\% to the total size of a prokaryotic genome \cite{Rath2015}, so it is hard to imagine that adding or removing a few spacers would affect the growth rate in a noticeable way.  Indeed, while there are various acknowledged sources of fitness cost for maintaining a CRISPR-Cas system \cite{Marraffini2008,Bondy-Denomy2014}, none of them significantly depends on the number of the CRISPR spacers \cite{Rath2015,Vale2015,Gophna2015}.  

Virtually all models of prokaryotic and viral coevolution driven by CRISPR immunity include some representation of the number of CRISPR spacers. In some models the array content is limited by a maximal number of spacers (see, for example, \cite{Childs2012}, where such  number is 8), or the number of spacers is determined dynamically as a result of competition between spacer acquisition and loss (such as in \cite{Iranzo2013, Bradde2017}). For a given set of environmental conditions, such as the abundance and variety of infecting viruses, the dynamic determination of the optimal number of spacers often manifests itself as dominance of bacterial subpopulation with such arrays. At the same time, the number of spacers plays a major role in determining the complexity of simulation because it is usually required to check all possible pairwise spacer-protospacer matches to determine the immune status of a pair of bacterial and viral strains. 

In this study, we propose a somewhat different view at the optimality of the number of spacers in CRISPR array. In particular, we ask a question of a rather idealized nature: What would be the number of spacers that maximizes protection of a given bacterium (rather than, for example, the survival of a bacterial species) from viruses? We show that the number of CRISPR spacers  is primarily limited by “dilution'' of CRISPR complexes carrying the most immune-active recently acquired spacers that target viral protospacers which had the least time to mutate. Our analysis requires a more detailed look at the kinetics of binding of CRISPR effector (a complex of Cas proteins with an individual protective CRISPR RNA, crRNA) to viral targets and distribution of crRNAs with particular spacers among the effectors. Since the origin and utility of the majority of spacers in each array are unknown, we made a simplifying assumption that all spacers in an array come from viral DNA and are used to repel viral infections. Another simplifying assumption we made is that instead of focusing on the actual evolution that occurs in ever-changing natural viral and bacterial communities, we compare the performance of arrays in their steady state for a given set of environmental parameters. We find that there exists a non-trivial optimal number of spacers, which maximizes the bacterial survival chances.

\section{THE MODEL}

\subsection{Basic assumptions}

\begin{figure}
	\centering
	\includegraphics[width=7in]{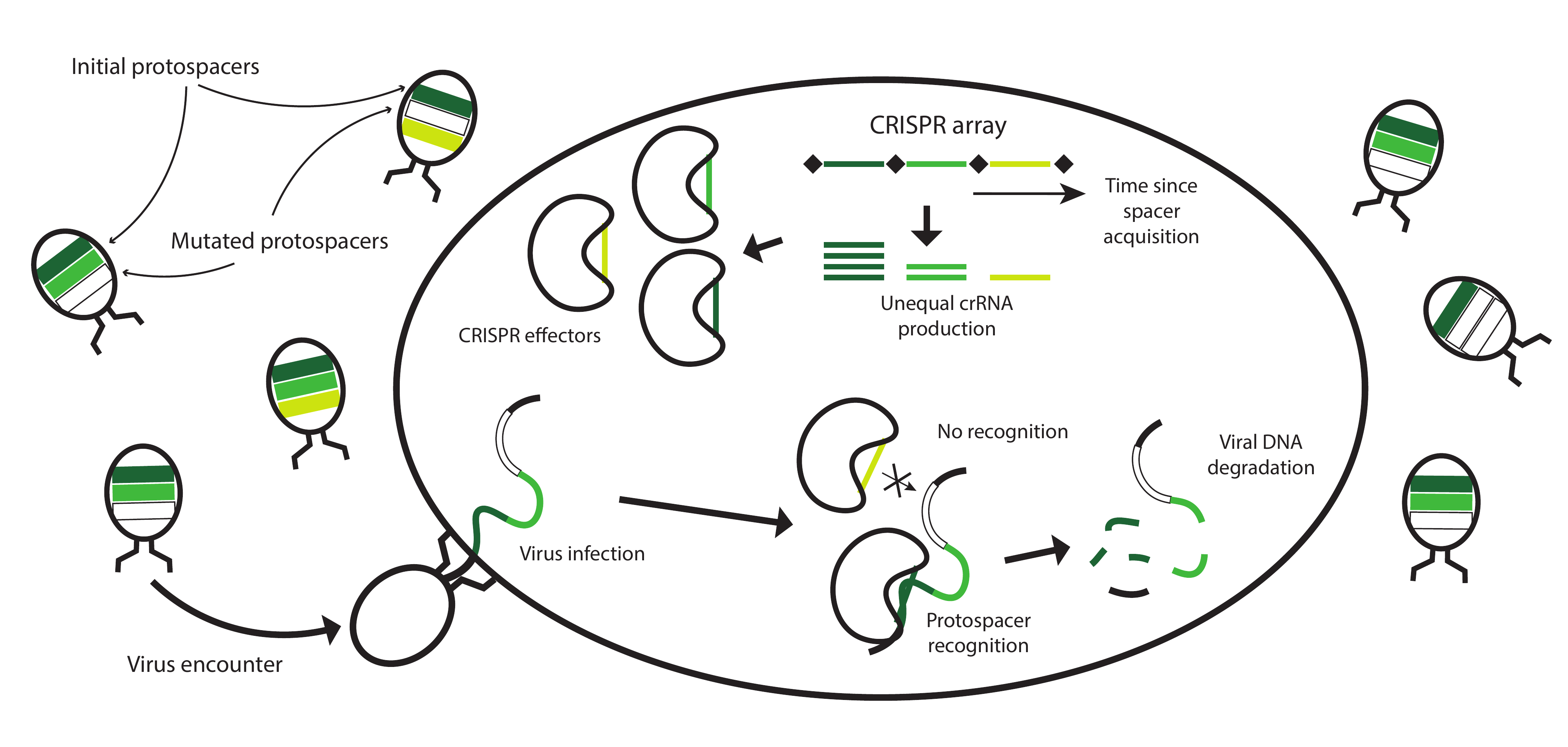}
	\caption{ \textbf{Functioning of CRISPR system.}
		Three spacers are colored according to their age from the time of their acquisition, from dark green marking the youngest spacers to yellow marking the oldest one. Phages carry protospacers colored similarly to their matching spacers; mutated protospacers are colored white. There are more mutated protospacers among older protospacers than among the younger ones.  Inside the cell, bean-shaped objects are CRISPR effector complexes armed with individual crRNAs. Complexes with younger spacers are more abundant than those with older ones. Viral DNA is shown to be simultaneously assessed by two CRISPR effector complexes: the dark green CRISPR spacer matches the non-mutated corresponding protospacer while the protospacer corresponding to the yellow spacer has mutated. The former interaction results in destruction of viral DNA while the latter leaves it intact.}
	\label{fig:model}
\end{figure}

Consider a prokaryotic cell with an active CRISPR-Cas system in a medium where phages capable of infection are present. The cell is attacked by individual viruses in a random and independent way: an attack is either repelled or kills the cell on a much shorter timescale than a typical time interval between subsequent attacks (Fig. \ref{fig:model}).  We assume that CRISPR-Cas immunity is the only protection available against the infection and each infection, which overcomes the CRISPR defense, results in cell death.

The CRISPR array consists of a number of spacers acquired during previous viral attacks and does not change over the timescale of analysis. Each spacer corresponds to a protospacer in DNA of viruses capable of  infection. A match between a spacer and a protospacer is a necessary (but not sufficient) condition for efficient defense from infection. Protospacers may mutate, making now partially complementary spacer ineffective. Thus, it could be beneficial for a cell to pick up more than one spacer from each virus thus reducing the probability of failure of CRISPR-Cas system to recognize viral DNA \cite{Levin2013}. This allows the cell to hedge against mutation in single protospacer leading to more reliable recognition of the virus and increased probability of survival. It is intuitively appealing to arm more CRISPR effectors with newer, more recently acquired spacers rather than with the older ones so that the corresponding protospacers would have had less time to mutate. The older the spacer, the higher is the probability that the next encountered virus will have a corresponding protospacer mutated leading to cell death. Indeed, there is a strong preference for spacers acquisition at one end of CRISPR array \cite{ Diez-Villasenor2013, Jackson2017}. As a result, spacers in natural arrays are ordered according to their age, with more recently acquired spacers located closer to promoter from which the array is transcribed. While the abundance of individual crRNAs is a complex function of their processing rate from pre-crRNA CRISPR-array transcripts and stability, promoter-proximal crRNAs are expected to be generally more abundant that promoter-distal ones. This effect is expected from transcription polarity and made more pronounced by the palindromic nature of CRISPR repeats, which should promote transcription termination by RNA polymerase. Thus comes the second element of selective pressure over the number of CRISPR spacers: A too long array will ``dilute'' the concentrations of CRISPR effector complexes armed with most recently acquired and thus most efficient spacers, replacing them with older spacers whose complimentary protospacers had a longer time to accumulate mutations. For simplicity, we assume that a single mismatch between a spacer and its protospacer makes the spacer ineffective \cite{Barrangou2007}. While the reality is more complex and certain mutations in a protospacer do not preclude its recognition by the spacer \cite{Semenova2011}, mutations in protospacer adjacent motif \cite{Fischer2012,Shah2013} or seed region \cite{Semenova2011}) indeed abolish CRISPR interference and it is mutations of this kind that we consider in our work.

The optimal number of spacers may be thought of as emerging from competition between the opposing ``more reliable recognition'' and ``dilution'' trends. We ignore the fitness cost of maintaining a CRISPR array, often considered to be consisting of two parts: spacer-number-independent and spacer-number-dependent \cite{Childs2012,Iranzo2013}. While duplication of CRISPR-Cas system DNA must have its cost, yet every new spacer constitutes a very small part of CRISPR-Cas DNA (which itself is a small part of cellular genome) and such cost is ignored. 

To summarize, we try to determine the optimal number of spacers in a CRISPR system illustrated in Fig.~\ref{fig:model} under the following simplifying assumptions:

\begin{itemize}
	\item The cutting of viral DNA is possible when there is a perfect match between the spacer and protospacer, and a single mismatch makes the spacer-protospacer pair useless for cell protection/CRISPR interference \cite{Semenova2011,Fischer2012,Shah2013}.
	\item  Probability for a CRISPR effector complex to contain crRNA with a particular spacer decreases exponentially with the age of the spacer.  
	\item The total number of CRISPR effector complexes in the cell is constant. While there is some evidence that cas genes expression might be regulated in vivo \cite{Pul2010},  the assumption that at given conditions expression levels are constant seems to be reasonable. 
	\item There is only a single copy of viral DNA inside the cell upon infection, i.e., the multiplicity of infections is low. 
	\item CRISPR arrays have a constant size and composition that do not change on the timescale of viral infection, i.e., there is no CRISPR adaptation.
	\item A single encounter between CRISPR-effector and virus resolves on a shorter timescale than the time between subsequent encounters.
	\item We do not take into account any fitness costs of maintaining an array of a given length \cite{Gophna2015,Vale2015}.
\end{itemize}

\subsection{Probability of interference}
Assume that a cell carries an array consisting of $S$ spacers which we number in the direction of age such that the most recently acquired spacer is assigned number 1.  The cell is being attacked by a virus and CRISPR defense comes into play.
The probability $B_i$ for CRISPR effector charged with crRNA with spacer $i$ to bind to the corresponding protospacer (or the fractional occupancy of the protospacer) is controlled by competition between binding and dissociation events which are described by the first and second terms in the right-hand
side of the following kinetic equation,
\begin{align}
\label{kinetic}
\frac{dB_i}{dt}=k^+(1-B_i) C_i - k^- B_i.
\end{align} 
Here $k^+$ and $k^-$ are the association and dissociation rate constants for a matching spacer-protospacer pair and $C_i$ is the copy number (uniquely related to its concentration since the volume of the cell is constant) of CRISPR effectors carrying the $i$th spacer crRNA. The steady state binding probability (or the fraction of time the corresponding protospacer is recognized by CRISPR effector) is 
\begin{align}
\label{steady}
B_i=\frac{k^+ C_i}{k^+  C_i + k^-}=\left[1+k^-/(k^+C_i)\right]^{-1}.
\end{align} 
Now we compute how $C$ CRISPR effectors present in the cell pick up
crRNAs with particular spacers. We have postulated that the number of
effector complexes that acquired spacer $i$ decreases exponentially
with the age of $i$. That is, each next spacer is $\delta$ times less
likely to be present in CRISPR effector complex than its younger
neighbor. We will further refer to $\delta$ as "crRNA decay
coefficient" since we assume that the exponential decrease in the
number of crRNA molecules with a defined spacer causes the
corresponding decrease in the number of CRISPR effector complexes with
this crRNA \cite{Zoephel2013}. % However, in vivo, $\d$ is certainly a much more complex
% function of spacer seniority.  [CITATIION NEEDED]
Hence the number of effector complexes $C_i$ with crRNA with spacer $i$ is
\begin{align}
\label{delta}
C_i=C_1 \delta^{i-1}.
\end{align} 
We determine $C_1$ from the condition that the total number of CRISPR effector complexes is $C$ by summing the corresponding geometric progression 
\begin{align}
\label{C}
C_i=C \delta^{i-1} \frac{1-\delta}{1-\delta^{S}}
\end{align} 
Substituting (\ref{C}) into (\ref{steady}) produces a complete expression for the binding probability between the $i$th spacer-protospacer pair,
\begin{align}
\label{B}
B_i=\left(1+ \frac{1}{\b}\frac{1}{\delta^{i-1}} \frac{1-\delta^{S}}{1-\delta}\right)^{-1}.
\end{align} 
Here $\b\equiv C k^+/(k^-)$ is the dimensionless coefficient which determines the ``binding efficiency'' of CRISPR effector. The larger $\b$, the larger fraction of time the effector spends bound to matching protospacer. The biological meaning of $\b$ becomes clear if one considers a CRISPR array consisting of a single spacer. Then the binding probability becomes the function of $\b$ only,
\begin{align}
\label{Bs}
B=\frac{1}{1+ 1/\b}.
\end{align} 
In such a case, the binding probability  depends on how $\b$ compares to 1: If $\b \gg 1$, the binding probability saturates to its maximum equal to 1, while if $\b \ll 1$, the binding probability becomes proportional to $\b$.  For $\b=1$ the binding probability is precisely 1/2. 

Assume that binding of every CRISPR effector to its matching protospacer proceeds independently of binding by other effectors to theirs, i.e., protospacers are well-separated in viral genomes.  The total rate of interference is then proportional to the sum of binding probabilities of matching spacer-protospacer pairs and the probability of survival of viral DNA $P(t)$ decays with a simple exponential kinetics,
\begin{align}
\label{dec}
\frac {dP(t)}{dt}=-a P(t) \sum_{i} B_i; \;\; 
P(t)=\exp\left(-a  t \sum_{i}B_i \right).
\end{align}
Here $a$ is the viral DNA degradation rate constant, which we consider to be a fixed property of a CRISPR-effector universal for all spacer-protospacer pairs. Hence the probability of successful interference
\begin{align}
\label{int1}
I=1-P(\t),
\end{align}
where $\t$ is the effective time of interference, roughly equal to the time of the duplication of viral DNA. In other words, for successful termination of infection, the CRISPR effector complexes have to destroy the viral DNA before or during the first round of its duplication. Destruction of individual viral genomes at later times can not prevent the runaway viral DNA replication and productive infection. Introducing a dimensionless parameter $\chi\equiv \t a$, which characterizes the interference efficiency, turns Eqs.~(\ref{int1} and \ref{B}) into
\begin{align}
\label{int}
I=1-\exp \left[-\chi \sum_{i}B_i\right]=\\
\nonumber
1-\exp \left[-\chi \sum_{i}\left(1+
\frac{1}{\b}\frac{1}{\delta^{i-1}} \frac{1-\delta^{S}} {1-\delta}\right)^{-1}  \right].
\end{align}
Constants $\b$ and $\chi$ have simple interpretations in terms of
familiar Michaelis kinetics: The process of interference can be viewed
as a transformation of viral DNA catalyzed by CRISPR effectors.  The
binding efficiency $\b$ corresponds to the inverse Michaelis constant
per substrate concentration, $S/K_M$ in standard notations,
(\ref{Bs},\ref{int}). The interference efficiency $\chi$ corresponds
to the maximal reaction rate in Michaelis kinetics, i.e. the rate at which viral DNA bound by CRISPR effector is degraded. 
\subsection{Survival probability}
\begin{figure}
	\centering
	\includegraphics[width=5in]{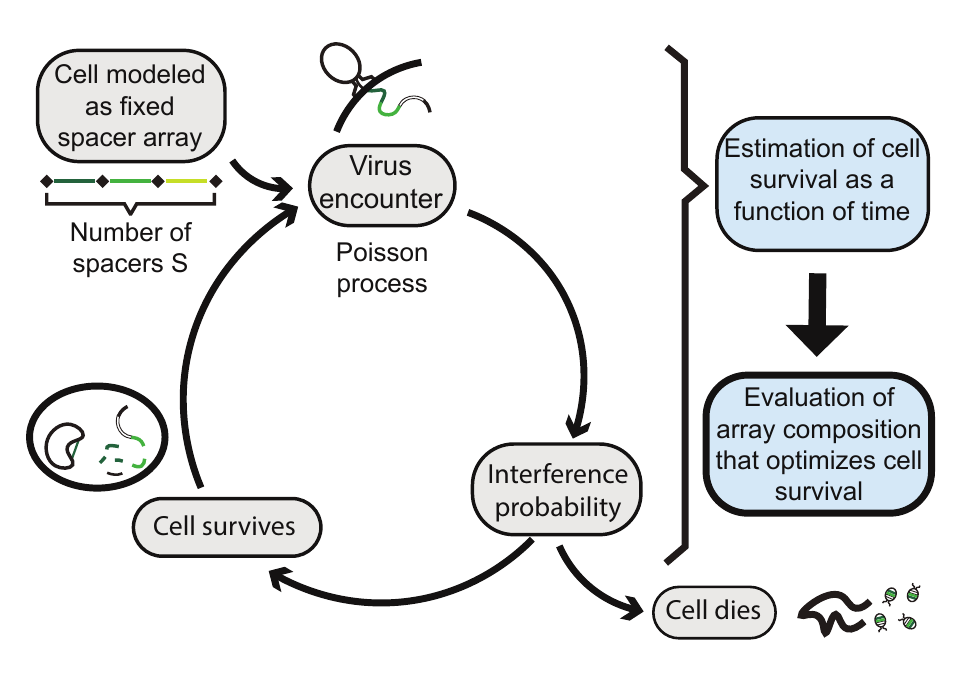}
	\caption{\textbf{Scheme of calculations.} A cell with $S=3$ CRISPR spacers encounters viruses as a Poisson process with an average rate $rN$. During each encounter there is either a successful interference with probability $I$ or the cell dies with probability $1-I$. We evaluate the probability $E(t)$ of the cell to survive till time $t$ as the measure of performance of its CRISPR-Cas system.}
	\label{fig:calc}
\end{figure}

Assume that a virus infecting a cell at a given moment is drawn from a big pool with a probability of infection proportional to the concentration of its type $v$ and that infections by different viruses are independent of each other. Then the probability $A_k$ to experience $k$ infections over time $t$ is given by a Poisson distribution with the average number of infections $rNt$ scaling linearly with time. 
\begin{align}
\label{poiss1}
A_k(t)=\frac{(r N t)^k }{k!}\exp (-r N t),
\end{align}
where $r$ is a proportionality coefficient considered to be the same for all viruses and $N$ is  concentration of the viral particles. To survive during a given time, each cell needs to repel all infections happening within this time, hence the probability of survival till time $t$ is 
\begin{align}
\label{poiss2}
\sum_{k=0}^{\infty}A_k(t) I^k = \exp [-r N t (1-I)].
\end{align}
Here $I$, defined in Eq.~(\ref{int}), is the probability to survive a single infection, i.e., the probability of successful CRISPR interference. From our assumption that viruses infect independently of each other it follows that the probability $E(t)$ for a cell to survive in the medium with several different viruses with concentrations $v_j$ is given by the product of survival probability determined for each virus separately,
\begin{align}
\label{poiss3}
E(t)= \prod_j \exp [-r N_j  t (1-I_{j})].
\end{align}
This is sketched in Fig.~\ref{fig:calc}. The probability of CRISPR interference with a single infection $I_j$ is defined as in (\ref{int}) with the sum running over all spacers taken from the $j$th virus. In the following we use $E(t)$ as the measure of overall CRISPR system performance. % – the overall measure of how good for the cell to have such CRISPR system.

\subsection {Single viral species}
To illustrate and further develop the general statement (\ref{poiss3}), consider a scenario of a single viral species infecting a cell that has a CRISPR array with just two spacers. The immunity depends on the mutation status of corresponding protospacers in viral population. In this model, the mutation status of the spacer will be defined as the fraction of mutated protospacers in the viral population. We denote by $m_1$ and $m_2$ the probabilities for the first and second protospacers to remain mutation-free and thus recognizable by CRISPR effectors. If the total concentration of viral particles is $N$ the concentration of the “wild type” variant without any mutations is $m_1 m_2 N$, the concentration of the variant with mutation in the second protospacer is $m_1(1-m_2) N$, the concentration of the variant with mutation in the first protospacer is $m_2(1-m_1) N$, and the concentration of the variant with mutations in both protospacers, i.e., an escape mutant not subject to CRISPR interference, is $(1-m_1)(1-m_2) N$. From Eqs.~(\ref{int} and \ref{poiss3}) and our assumption that a mutation in protospacer renders the corresponding spacer completely inefficient,  it follows that the survival probability in such case is 
\begin{align}
\label{surv2}
E(t)= 
\exp \left(  - r N t \left\{
m_1m_2 \exp [-\chi (B_1+B_2) ] +\right. \right. \\
\nonumber
\left. \left. +m_1(1-m_2) \exp [-\chi B_1 ] +
m_2(1-m_1)\exp [-\chi B_2 ] -
(1-m_1)(1-m_2)\right\} \right)
\end{align}
The last term in the exponent corresponds to the probability to experience no infection by viruses with both mutated protospacers (in which case $I_{4}=0$ since such an infection would result in cell death).  Transforming the expression in the exponent, we obtain
\begin{align}
\label{surv3}
E(t)= \exp \left[  - r N t\left( 
\prod_{i=1}^2\left\{1-m_i[1-\exp (-\chi B_i)]\right\}\right) \right ].
\end{align}
This expression has a simple probabilistic interpretation: The $i$th term in curly brackets describes the probability of failure of CRISPR effector complexes armed with the $i$th spacer crRNA. The product of such terms describes the probability of failure of all CRISPR effectors and thus the death of the cell. The expression (\ref{surv3}) is the probability for the Poisson process of “failures” of CRISPR system to have zero counts or no failures at all, which translates into survival of the cell. Mutual independence of encounters with different mutation variants of the virus simplifies the survival probability of the cell to the probability of not to be affected by the “average'' encounter repeated $rNt$ times. This simple interpretation allows us to generalize (\ref{surv3}) to cases of arrays containing more than 2 spacers, replacing the upper limit of the product by an actual number of CRISPR spacers $S$,
\begin{align}
\label{surv4}
E(t)= \exp \left[  - r N t \left( 
\prod_{i=1}^S\left\{1-m_i[1-\exp (-\chi B_i)]\right\}\right) \right ].
\end{align}
To reduce the number of independent parameters in Eq.~(\ref{surv4}) and in the following expressions for the survival probability, we estimate $m_i$. We assume that spacers were acquired to the array in a periodic fashion, that is, the time intervals $t_{ins}$ between subsequent acquisition of spacers were the same. The probability for a protospacer to remain mutation-free decreases exponentially with time, and the ``age'' of the $i$th protospacer is proportional to $i$. Hence, the probability of a perfect match for the $i$th spacer-protospacer pair at the middle of the time interval between spacer acquisitions can be approximated as $\m^{i-1/2}$. Here $0<\m<1$ is the probability for a protospacer in viral DNA not to undergo any mutations during $t_{ins}$ and $-1/2$ in the exponent stands for assessing the cell survival probability in $t_{ins}/2$ time units after the acquisition of the last spacer, i.e. in the middle of the interval between spacer acquisitions. The parameter $\m$ depends on genetic and environmental factors such as the rate of mutations in viral DNA, the size of the viral population, the size of protospacer, and the average rate at which cells acquire new spacers.
Eq.~ (\ref{surv5}),
\begin{align}
\label{surv5}
E(t)= \exp \left[  - r N t \left( 
\prod_{i=1}^S\left\{1-\m^{i-1/2}
[1-\exp (-\chi B_i)]\right\}\right) \right ],
\end{align}
together with the binding probability (\ref{B}), completely define the
survival probability of a cell with a given number of spacers $S$ as a
function of dimensionless parameters $\mu$, $\chi$, $\d$ and
$\b$. Note that the optimal number of spacers does not depend on the
total time of observation  $t$ that was used for cell survival evaluation: In Eq.~(\ref{surv5})  the position of the maximum of $E(t)$ is determined by the maximum of the product in the exponent and is independent of $rNt$.

\begin{figure}
	\centering
	\includegraphics[width=6in]{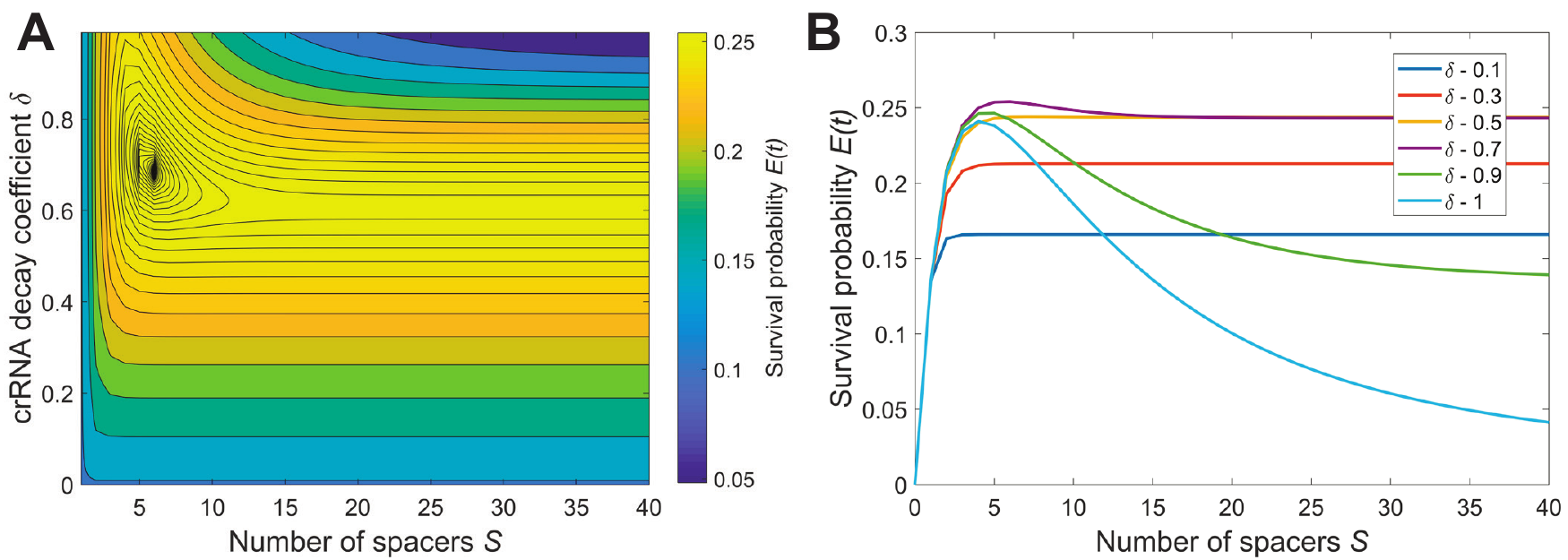}
	\caption{\textbf{Typical survival probability profile.} (A)  Plot of survival probability $E(t)$ vs.
		the crRNA decay coefficient $\d$ and the number of spacers in CRISPR array $S$. Other parameters are: $\b=1$,  $\chi=2$,  $\m=0.9$, and $rNt=5$. (B) Six curves of $E(t)$ vs. $S$ for various values of $\d$ and same
		$\b$, $\chi$, $m$, and $rNt$ as in the panel A.}
	\label{fig:typical}
\end{figure}

A typical dependence of survival probability $E(t)$  on the crRNA
decay coefficient $\d$ and the number of spacers $S$ is shown in
Fig.~\ref{fig:typical}. We inferred the interference probability
$I_1\approx 0.7$ of a single spacer array from the experimental data
\cite{Strotskaya2017} and set the binding efficiency $\b=1$ and the
interference efficiency $\chi=2$ to reproduce the measured
single-spacer interference probability. The probability for a
protospacer not to mutate over the typical period between spacer
acquisition was chosen to be $\m=0.9$. The typical number of
infections over the time of observation was $rNt = 5$. It follows from Fig.~\ref{fig:typical} that the survival is maximized for $\d \approx 0.7$ and $S=6$. In panel B the dependence of $E(t)$ vs. $S$ is shown for several values of $d$. Curiously, for low $d$, the survival $E(t)$ does not noticeably decrease for large $S$. It happens because of the exponential suppression in frequencies of crRNA with older spacers in effector complexes: no matter how long the array is, only crRNA with the first few spacers are mainly used by effectors. Thus, an ``automatic'' cutoff is implemented.

\begin{figure}
	\centering
	\includegraphics[width=5.5in]{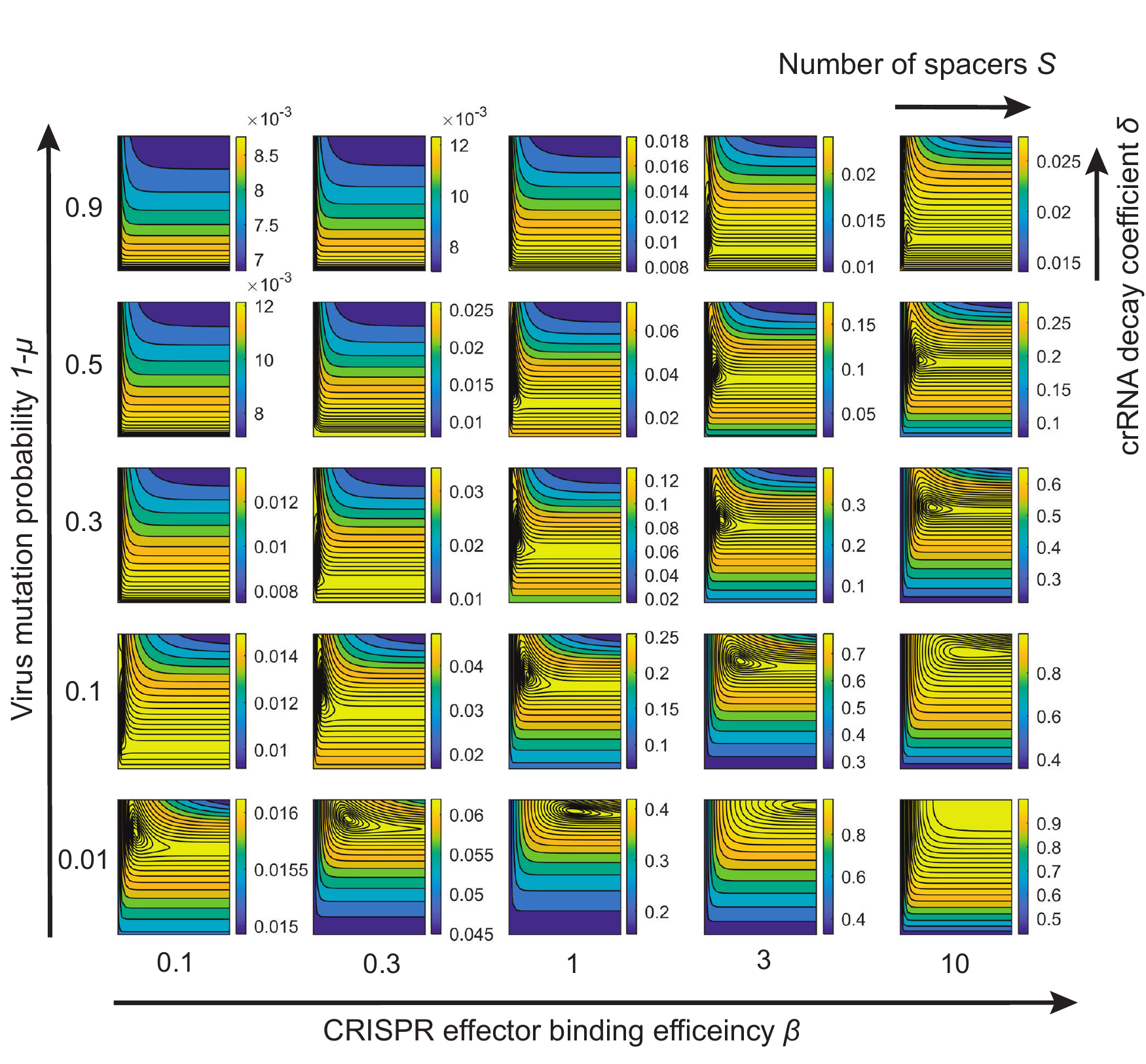}
	\caption{\textbf{Effects of mutation rate and binding efficiency.} A set of 25 panels                   illustrating how the survival probability depends on $S$ and $\d$ for various values of protospacer mutation probability  $1-\m$ and binding efficiency of effectors $\b$. The $\d$ and $S$ axes in each small panel have the same range as in the panel A in Fig.~\ref{fig:typical}, while the scale of the heat-map varies  and is indicated to the right of each panel. The external axes describe the variation of mutation probability $1-\m$ and effector binding efficiency $\b$.  In all panels $\chi=2$ and $rNt=5$.}
	\label{fig:results}
\end{figure}

Naturally, the optimal number of spacers depends on such parameters as protospacer mutation rate $1-\mu$ and the efficiency of effector binding to its targets $\b$: In Fig. ~\ref{fig:results} we show how the plot of the “typical case'' shown above in Fig.~\ref{fig:typical} is affected by changes in these system parameters. An increase in the mutation rate shifts the optimum towards fewer spacers or stronger reliance of the CRISPR-Cas system on crRNA with the first spacer. In the extreme case this can lead to the optimal array containing one spacer only (Fig. ~\ref{fig:results}, top-left corner). This corresponds to the case when there is a very high chance that older spacers have mutated, so the benefit from using the second spacer cannot overcome the decrease in the number of effector complexes loaded with crRNA containing the first, most recently acquired spacer. In contrast, an increase of CRISPR interference efficiency shifts the optimum towards more CRISPR spacers and more equal contribution of spacers of different age (Fig. ~\ref{fig:results}, bottom-right corner). An increase in the binding efficiency  leads to a larger fraction of time the effector spends bound to the protospacer ultimately leading to binding  saturation. In this case the sharing of CRISPR effectors between crRNAs with different spacers is beneficial as it allows the effectors to reduce competition for the same protospacer. An increase in the CRISPR interference efficiency $\chi$ also leads to an increase in survival probability (data not shown).

\begin{figure}
	\centering
	\includegraphics[width=6.5in]{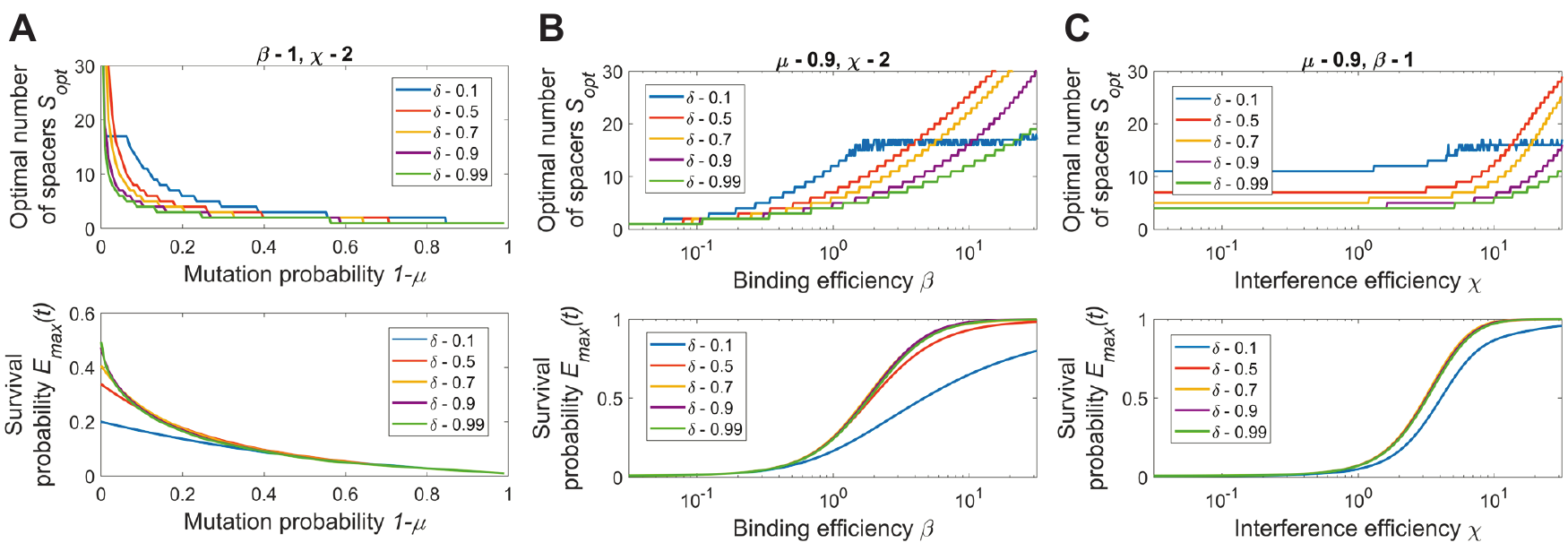}
	\caption{\textbf{Effect of parameters on the optimal number of spacers and the maximal survival probability.} The optimal number of spacers and corresponding survival probability as functions of one of the array-unrelated parameters: (A) As function of mutation probability $1-\mu$, other parameters are $\beta = 1$ and $\chi = 2$. (B) As function of binding efficiency $\beta$, other parameters are $\mu = 0.9$ and $\chi = 2$. (C) As function of interference efficiency $\chi$, other parameters $\mu = 0.9$ and $\beta = 1$. The average number of viral infections was $rNt=5$ in all panels.}
	\label{fig:optimum_change}
\end{figure}

For a more detailed study of the optimal number of spacers, we conducted the following calculations: for each set of “array-independent'' parameters $\m, \b, \chi$ we analyzed the CRISPR efficiency in the whole range of the number of spacers $S$ and crRNA decay coefficients $\d$. The number of spacers $S_{opt}$ and crRNA decay coefficient $\d_{opt}$ that maximized survival probability, as well as the maximal survival probability itself $E_{max}(t)$ are plotted in Fig. ~\ref{fig:optimum_change}. As discussed above, higher viral mutation rates lead to lower survival probability and fewer spacers (Fig. ~\ref{fig:optimum_change}A). For very high mutation probability (above 0.7) the CRISPR interference efficiency approaches zero for all values of other parameters. The mutation rate of viruses caps the CRISPR efficiency as the probability to survive the infection is constrained by the probability $I_{max}$ that at least one of viral protospacers has not mutated.
\begin{align}
\label{cap}
I_{max} = 1-\prod_{i=1}^{S}(1-\mu)^{i-1/2}
\end{align}

\begin{figure}
	\centering
	\includegraphics[width=6in]{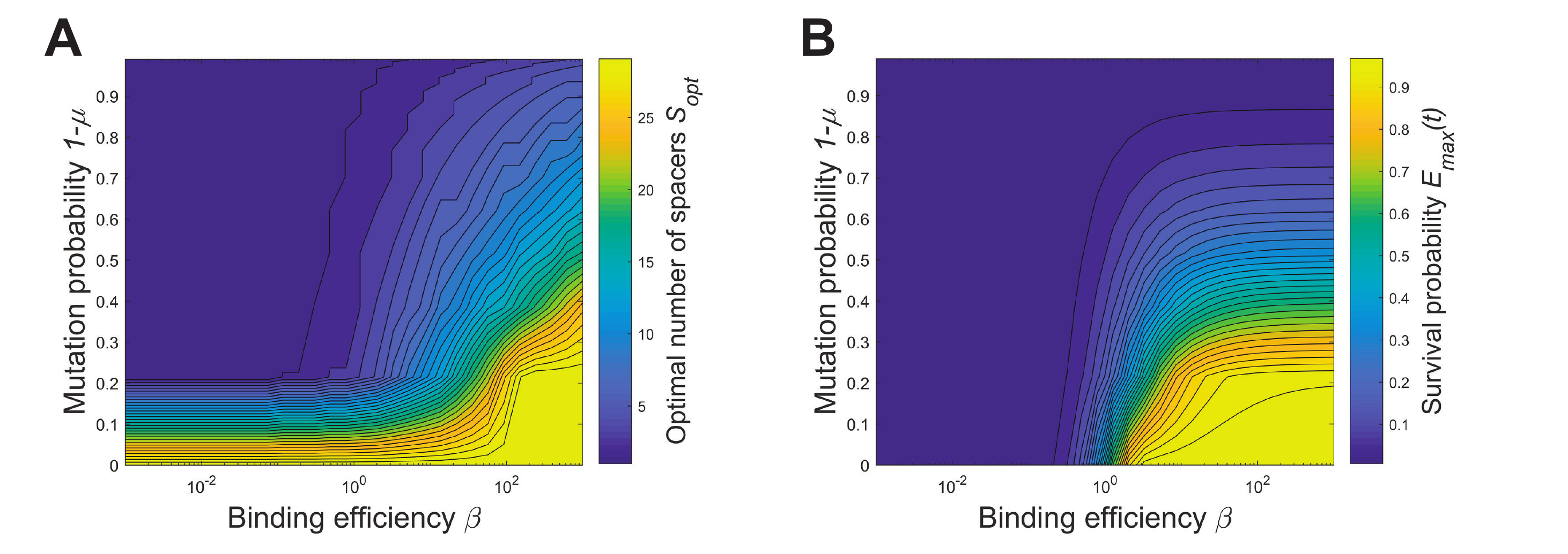}
	\caption{\textbf{The optimal number of spacers and  maximal
            cell survival probability.} The optimal number of spacers
          (A) and the maximal cell survival probability (B) are shown
          vs. a range of binding efficiencies $\beta$ and mutation
          probabilities $1-\mu$ for $rNt=5$ and $\chi=2$.} 
	\label{fig:results_optimum}
\end{figure}

On the other hand, a high binding $\b$ or interference efficiency $\chi$ lead to arrays with more spacers and higher survival probability (Fig. ~\ref{fig:optimum_change}B, C). In this case, more CRISPR effectors can complex with crRNAs with older spacers without interfering with the binding to crRNAs with younger spacers due to the system saturation. Arrays with more spacers both increase the viral DNA degradation rate and, more importantly, reduce the chance that the cell becomes unprotected if some of protospacers mutate. This suggests a correlation between the optimal number of spacers $S_{opt}$ and the maximal protective performance of CRISPR-Cas system $E_{max}(t)$. Comparing the optimal number of spacers and maximal survival probability heat-maps shown in Fig. ~\ref{fig:results_optimum}, one sees that the parameters that produce high survival probability indeed correspond to arrays with  relatively  many (more than 10 spacers) spacers. 

Figs.~\ref{fig:optimum_change} and \ref{fig:results_optimum} lead to a conclusion that there is a definite set of parameters for which CRISPR-Cas systems are efficient. The virus mutation probability should remain low on the timescale of spacer acquisition, while the binding of effector complexes to target protospacers and the rate of degradation of viral DNA should be high. This set of parameters favors arrays with more spacers. It implies a simple rule: the array can contain many spacers  and be efficient or contain few spacers  and be inefficient. In reality, the array composition could change on the timescale of viral infections, which may increase CRISPR interference efficiency.  This, however, goes beyond the important assumption of our model of the static nature of the array and thus is beyond our present consideration. On the other hand, it shreds the light on the adaptive immunity as the only efficient way of CRISPR-related defense in the viral environments with fast mutation rates.

\subsection{Multiple viral species}
Consider now a more realistic scenario of a cell confronting several distinct viral species. Using the same logic as in the section above and, specifically considering infections by different viruses being independent of each other, we conclude that the survival probability is given by the Eq.~(\ref{poiss3}), where the index of the product $j$ enumerates all viral species, including their mutation variants, present in the system. The interference term associated with a viral species $j$ not targeted by any spacer present in a given array is zero,  $I_{j}=0$. The corresponding term in the survival probability $\exp (-rNt v_j) $ describes the probability for a cell not to encounter such a virus till time $t$.  

Similarly to the case of single viral species, we account for mutation variants of each virus and reduce (\ref{poiss3}) to the product running over only distinct viral species. In order to simplify further analysis we denote by $v_i$ the fraction of the $i$th virus in the total number of viruses $N$ so that $v_i = N_i/N$, where $N_i$ is the number of viral particles of species $i$. This results in the following expression for survival probability of a cell with a given combination of spacers, 
\begin{align}
\label{surv_tot}
E_{c}(t)= \exp \left[  -rNt \sum_{j=1}^{\nu} v_j \left( 
\prod_{i \in \{S_j\}}\left\{1-m_i[1-\exp (-\chi B_i)]\right\}\right) \right ].
\end{align}
Here the sum over $j$ counts all $\nu$ viral species while the product over $i$ enumerates all spacers $\{S_j\}$ taken from the $j$th virus. As in (\ref{surv4}), we approximate $m_i$ by $\m^{i-1/2}$ assuming again that spacers are acquired in a periodic fashion, with equal times between acquisitions.

The equation (\ref{surv_tot}) describes survival probability of a cell with a given CRISPR array characterized by sets of spacers $\{S_j\}$ taken from viral species $j$. In order to evaluate the overall performance of a CRISPR array with  $S$ spacers, we need to enumerate survival probabilities for all combination of spacers in such an array. 
To do so, we assume that the probability to acquire a spacer from a given viral species is proportional to the fraction of such species in the total viral pool. Hence the probability of an array to have certain combination of spacers is  
\begin{align}
P_{c} = \prod_{k=1}^S v_k,
\label{prob}
\end{align}
where $v_k$ is the relative concentration of viral species from which the spacer $k$ has been acquired. For example, an array of two spacers $(a,b)$ in a system populated by two viral species 1 and 2 with relative concentrations $v_1$ and $v_2$ can be in any of the following four forms with corresponding probabilities: $P_{(1,1)}=v_1^2$, $P_{(1,2)}=P_{(2,1)}=v_1v_2$, and $P_{(2,2)}=v_2^2$. 

The average survival probability of a cell in a multiviral medium is a sum of survival probabilities corresponding to each combination of spacers $E_c$, weighted by the probability to acquire such a combination $P_c$, and the summation runs over all combinations of spacers.
\begin{align}
\label{surv_by_comb}
E(t)= \sum_{c} E_{c}(t) P_{c}.
\end{align}

\begin{figure}
	\centering
	\includegraphics[width=4.5in]{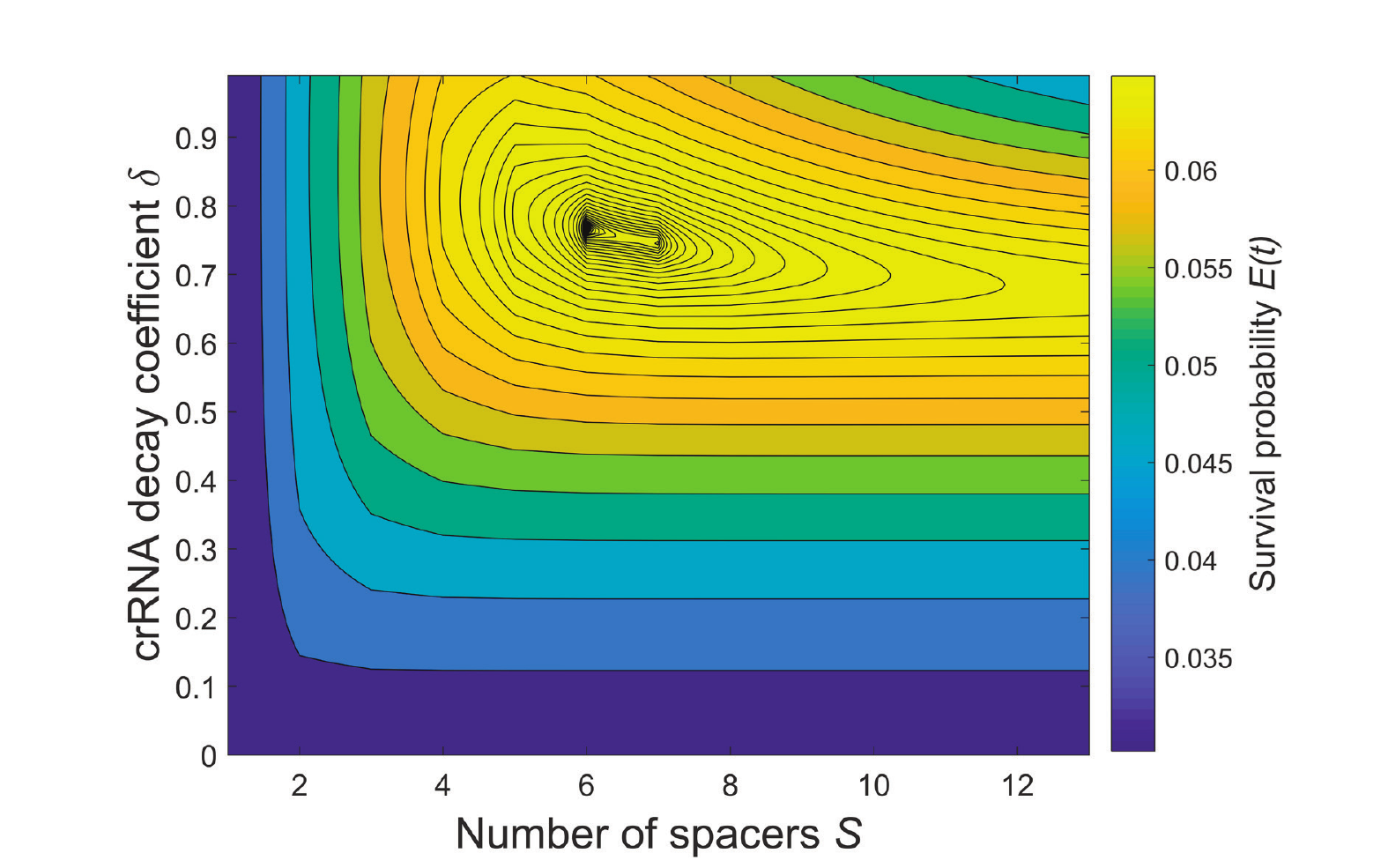}
	\caption{\textbf{CRISPR performance for two virus species.} Plot of the survival probability $E(t)$ as a function of crRNA decay coefficient $\d$ and the number of spacers $S$ of a cell confronting  two different viruses with equal population sizes, $\nu_1 = \nu_2 = 0.5$. The binding efficiency is $\beta=1$ and the interference efficiency is $\chi=2$. Viral mutation probability $1-\mu$ is equal to 0.1 and $r Nt=5$.}
	\label{fig:typical_multi}
\end{figure}

A typical plot of $E(t)$ is presented in
Fig.~\ref{fig:typical_multi}. In this calculation we considered two
species of viruses with the same population size $v_1 = v_2 =
0.5$. The values of other parameters were the same as in
Fig.~\ref{fig:typical}: The binding efficiency $\b=1$, the
interference efficiency $\chi=2$, the probability for a protospacer
not to mutate over the typical period between spacer acquisition
$\m=0.9$, and the typical virus encounter number $rNt=5$. Comparing to
the single-virus case in Fig.~\ref{fig:typical}, the total number of
viral particles is the same, but the virus pool is  now split between two species.

In general, the shape of the survival probability $E(t)$ profile is similar to the single-virus case and $E(t)$ reaches its maximum for certain $\d$ and $S$. However, comparing the optimal number of spacers, crRNA decay coefficient, and survival probabilities between the single- and two-virus cases (Figs.~\ref{fig:typical}A and \ref{fig:typical_multi}), one sees that in the two-virus case the maximum is generally shifted towards  arrays with more spacers, and $E(t)$ is lower. For a given set of parameters the addition of the second virus does not significantly shift the optimal $S$ and $\d$ but drops the survival probability dramatically. If the virus mutation rate is lower and the CRISPR interference efficiency is higher, the presence of an additional viral species will affect the optimal $S$ and $\d$ more strongly. However, relating the model parameters to the experimental results \cite{Strotskaya2017}, it is unlikely that the CRISPR efficiency in the multivirus environment can be significantly higher in vivo than the numbers shown in  Fig.~\ref{fig:typical_multi}.

\begin{figure}
	\centering
	\includegraphics[width=7in]{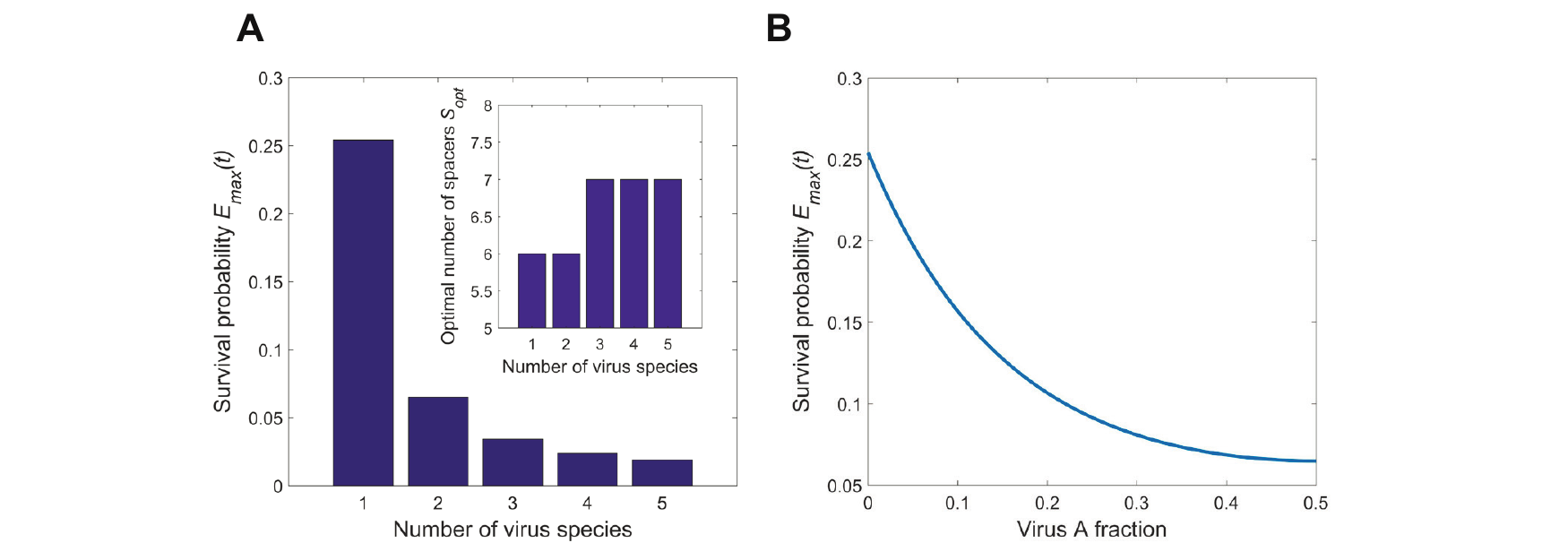}
	\caption{\textbf{Survival probability vs diversity of the
            virus pool.} Plots of the optimized over $\d$ and $S$ cell
          survival probability  and  the number of spacers vs the
          number of viral species and the composition of a
          two-virus pool for $\beta=1$,
          $\chi=2$, $\mu=0.9$  and $rNt=5$. (A) Maximal survival probability $E(t)$ (outer plot) and optimal number of spacers $S_{opt}$ (inner plot) as a function of the number of virus species $n$. The  abundance of virions belonging to different species in the viral pool are the same for all species, $\nu_1 =...= \nu_n = 1/n$. (B) The maximal survival probability vs the relative abundance of one of the viruses in a two-virus pool.}
	\label{fig:analysis_multi}
\end{figure}

When the number of virus species in the total virus pool increases even without a change in the total viral particles concentration, the survival probability approaches zero (Fig. ~\ref{fig:analysis_multi}A). This occurs because the efficient number of spacers is limited by the virus mutation rate and the number of effector complexes present in the cell (encoded in the coefficient $\b$). In other words, further increase in the number of spacers does not lead to any increase in protective function of CRISPR-Cas. Since an array of an effectively limited number of spacers has to contain spacers from more virus species, fewer spacers match each virus and the survival probability decreases.

Another observation is obtained considering the two-virus case and changing the ratio of those viruses in the pool (Fig. ~\ref{fig:analysis_multi}B). As expected, the survival probability reaches a maximum when the fraction of one virus approaches zero (which correspond to the single-virus case) and goes to a minimum when the two viruses are equally abundant.

This brings us to the conclusion that survival probability of a cell dramatically depends on the diversity of the viral pool.

\section{Discussions and Conclusions}
The function of CRISPR-Cas as prokaryotic adaptive immune system has
been extensively studied from the point of view of molecular
mechanisms.  Its ecological role and its contribution to the
"arms race" between prokaryotes and their viruses have been analyzed in
many evolutionary dynamics models and found to be very
complex  and often unpredictable. In this work, we qualitatively explored the forces affecting the number of spacers in a CRISPR array. We found that more spacers in a CRISPR array targeting a virus decrease the chances of the virus to escape detection through simultaneous mutation in all targeted protospacers.  Also, more spacers lead to more effective use of CRISPR effectors, distributing them between a larger number of target protospacers, which results in higher probability of viral DNA destruction. However, at the same time, more diverse crRNA repertoire results in fewer effector complexes charged with crRNAs containing recently acquired spacers that target protospacers least likely to mutate. The interplay of these forces leads to the optimum in the number of spacers per array, determined by the properties of the CRISPR system and viral environment in the following way: A better binding of the CRISPRS effectors to their targets and faster rate of target DNA degradation allow a bacterium to maintain more spacers in the array and increase its survival probability. Also, less frequent mutations in viral protospacers create an opportunity for hedging against those mutations by keeping more of previously acquired spacers. In contrast, a less efficient kinetics of binding and viral DNA cutting and faster-mutating viruses make arrays with fewer spacers more advantageous. A few comments on applicability of our results and biological insights that can follow from them are in order:

\subsection {Effects of dynamics and environment.}
Our results were derived explicitly assuming a steady state of the
CRISPR-virus dynamics. However, in previous research, both modeling
and experimental, it was shown that CRISPR systems are far from being
stable, undergoing periodic and irregular variations that play an
important role in their function
\cite{Berezovskaya2014,Childs2012}. While in our analysis we assumed
that the viral environment is constant (except for appearance of
mutant protospacers), the actual viral dynamics may affect the optimal
lengths of CRISPR arrays. Changes in the environment could explain an
increased efficiency of the shorter array in experimental condition
comparing to the wild strain \cite{Rao2017}. Moreover, the primed
acquisition of spacers could happen on the timescale of a virus attack
\cite{Semenova2016}, which would invalidate our basic assumption of
separation of timescales of viral attacks and spacer
acquisition. These factors, not analyzed in our work, could affect the optimal number of CRISPR spacers and are subject to further analysis.
\subsection {Comparison with existing results}
Our results generally agree with the main findings of models existing
in the field: We confirm that a higher diversity of viruses in the
environment results in a dominance of viruses over the CRISPR system
\cite{Iranzo2013, Weinberger2012}. This effect could be achieved by
either a high number of virus species in the environment or a high
mutation rate of viruses belonging to the single species (often
associated with large viral population). However, here we have also
shown that a diversity of virus species leads to arrays with more
spacers while a higher viral mutation rate leads to arrays with fewer
spacers. This agrees with a proposed hypothesis that a lower viral mutation rate leads to arrays with on average more spacers in thermophilic bacteria \cite{Weinberger2012}. 
Another important note on comparing our model with existing ones is related to the definition of probability of CRISPR immunity failure. Some of the models used a binary approach to immunity failure \cite{Childs2012}. Either the infected cell kills the virus or the virus kills the cell and reproduces normally. We define the CRISPR failure probability $1-I$ as the probability of viral DNA not getting cut by CRISPR effectors/executors during viral DNA duplication cycle. Distinguishing between these two approaches is important as it affects the interpretation of parameters obtained from experiments. For example, a CRISPR-Cas system can remain active in doomed or dead cells, resulting in lower viral burst size and fewer secondary infections \cite{Strotskaya2017}.  Our analysis based on \cite{Strotskaya2017} resulted in the estimate of the CRISPR failure probability around 30\% compared to $10^{-5}$ in \cite{Childs2012}.

\subsection {Importance of hairpins.}

One of important observations is that the equipartition of crRNA between CRISPR effector complexes is not optimal and a decrease of the fraction of older crRNA bound to effectors increases the overall efficiency of the immune response. While there is a limited pool of effectors, they serve better when binding to crRNAs with most recently acquired spacers. Since the probability that a spacer no longer matches the protospacer increases with time, Cas effectors should either have a higher affinity towards crRNA from younger spacers (which is impossible to accomplish) or crRNA containing more recent spacers should be more abundant. This latter may be implemented naturally owing to formation of hairpin by CRISPR repeats in the primary array transcripts \cite{Kunin2007,Brouns1993}. It is well known that hairpins have a potential to pause or terminate transcription elongation \cite{Wilson1995,Farnham1981}. The longer the array is, the more hairpins need to be transcribed and the higher the chance is that transcription would be terminated before the  RNA polymerase reaches the end of the array. This could result in more abundant shorter pre-crRNAs that include only the younger spacers. 

\subsection { Fitness cost of CRISPR system}

While in our study we ignored the fitness costs of an active CRISPR system, we find it important to discuss it as these were studied in various experimental works and included in some models \cite{Han2017}. It has been shown in a number of publications that the activity of CRISPR systems is under strong evolutional pressure. There are various factors that can contribute to the cost of CRISPR including genomic burden \cite{Kuo2009}, the cost of maintenance of cas genes \cite{Vale2015}, self-immunity \cite{Vercoe2013} and blockage of beneficial horizontal gene transfer (HGT) \cite{Marraffini2008}. However genomic burden seems not to be significant in most cases as even the largest of the CRISPR systems contribute only 1\% to the total size of a prokaryotic genome \cite{Rath2015}. In the case of self-immunity, it seems to be related to the very process of acquisition of new spacers, thus, self-immunity only indirectly affects the length of the spacer array \cite{Wei2015,Levy2015,Yosef2012}. For the cost of gene maintenance \cite{Vale2015} and blockage of HGT \cite{Gophna2015}, it has been shown that an increase in the number of spacers also does not have any significant fitness cost. Thus, in this work, we considered that the fitness cost of CRISPR system did not affect the optimal number of spacers in CRISPR array. In other words, there is no additional fixed cost of the spacer apart from the one arising from Cas effector dilution. That resulted in separation of the number of spacers question from the overall fitness. The factors described in this work affect the optimum length of the CRISPR array and the total fitness benefit of CRISPR system. And this total fitness benefit now can be compared to the fitness cost of CRISPR-Cas system maintenance, that will give the answer whether the CRISPR system will be effective or tends to be knocked out \cite{Jiang2013}.

\section{Conclusions}

\begin{itemize}
	\item We theoretically predict the optimal number of spacers in a
	CRISPR array which falls into reasonable range from the
	viewpoint of current experimental data and show that it
	depends on the interference efficiency of
	CRISPR effector, crRNA spacer-protospacer binding
	efficiency, and virus mutation rate.
	\item “Good” (from the “point of view” of the cell) conditions, such as high interference and
	binding efficiencies and slow mutation of viral
	protospacers favor arrays with more spacers. Conversely, less favorable conditions
	shift the optimum to arrays with fewer spacers. 
	\item Arrays containing only a few (less than 10) spacers offer
	significantly less
	protection from viral attacks.
	\item The majority of optimal array configurations have a non-uniform
	distribution of unique crRNAs among CRISPR effector complexes with a
	preference for crRNAs with more recently acquired spacers.
	\item Fighting against multiple viral species shifts
	the optimum towards arrays with more spacers and dramatically
	decreases the maximum efficiency of the  CRISPR system.
\end{itemize}

\section{Acknowledgments}

Y.I. was supported by FONDECYT (Chile) grant 1151524 and by SkolTech during his stay there.

% Either type in your references using
% \begin{thebibliography}{}
% Text
% \end{thebibliography}
%
% or
%
% Compile your BiBTeX database using our plos2015.bst
% style file and paste the contents of your .bbl file
% here. See http://journals.plos.org/plosone/s/latex for 

\bibliography{CRISPR}

\begin{thebibliography}{46}
\expandafter\ifx\csname natexlab\endcsname\relax\def\natexlab#1{#1}\fi
\expandafter\ifx\csname bibnamefont\endcsname\relax
  \def\bibnamefont#1{#1}\fi
\expandafter\ifx\csname bibfnamefont\endcsname\relax
  \def\bibfnamefont#1{#1}\fi
\expandafter\ifx\csname citenamefont\endcsname\relax
  \def\citenamefont#1{#1}\fi
\expandafter\ifx\csname url\endcsname\relax
  \def\url#1{\texttt{#1}}\fi
\expandafter\ifx\csname urlprefix\endcsname\relax\def\urlprefix{URL }\fi
\providecommand{\bibinfo}[2]{#2}
\providecommand{\eprint}[2][]{\url{#2}}

\bibitem[{\citenamefont{Makarova et~al.}(2006)\citenamefont{Makarova, Grishin,
  Shabalina, Wolf, and Koonin}}]{Makarova2006}
\bibinfo{author}{\bibfnamefont{K.~S.} \bibnamefont{Makarova}},
  \bibinfo{author}{\bibfnamefont{N.~V.} \bibnamefont{Grishin}},
  \bibinfo{author}{\bibfnamefont{S.~A.} \bibnamefont{Shabalina}},
  \bibinfo{author}{\bibfnamefont{Y.~I.} \bibnamefont{Wolf}}, \bibnamefont{and}
  \bibinfo{author}{\bibfnamefont{E.~V.} \bibnamefont{Koonin}},
  \bibinfo{journal}{Biology direct} \textbf{\bibinfo{volume}{1}},
  \bibinfo{pages}{7} (\bibinfo{year}{2006}), ISSN \bibinfo{issn}{1745-6150},
  \urlprefix\url{http://www.biology-direct.com/content/1/1/7}.

\bibitem[{\citenamefont{Bolotin et~al.}(2005)\citenamefont{Bolotin, Quinquis,
  Sorokin, and {Dusko Ehrlich}}}]{Bolotin2005}
\bibinfo{author}{\bibfnamefont{A.}~\bibnamefont{Bolotin}},
  \bibinfo{author}{\bibfnamefont{B.}~\bibnamefont{Quinquis}},
  \bibinfo{author}{\bibfnamefont{A.}~\bibnamefont{Sorokin}}, \bibnamefont{and}
  \bibinfo{author}{\bibfnamefont{S.}~\bibnamefont{{Dusko Ehrlich}}},
  \bibinfo{journal}{Microbiology} \textbf{\bibinfo{volume}{151}},
  \bibinfo{pages}{2551} (\bibinfo{year}{2005}), ISSN \bibinfo{issn}{13500872}.

\bibitem[{\citenamefont{Barrangou et~al.}(2007)\citenamefont{Barrangou,
  Fremaux, Deveau, Richards, Boyaval, Moineau, Romero, and
  Horvath}}]{Barrangou2007}
\bibinfo{author}{\bibfnamefont{R.}~\bibnamefont{Barrangou}},
  \bibinfo{author}{\bibfnamefont{C.}~\bibnamefont{Fremaux}},
  \bibinfo{author}{\bibfnamefont{H.}~\bibnamefont{Deveau}},
  \bibinfo{author}{\bibfnamefont{M.}~\bibnamefont{Richards}},
  \bibinfo{author}{\bibfnamefont{P.}~\bibnamefont{Boyaval}},
  \bibinfo{author}{\bibfnamefont{S.}~\bibnamefont{Moineau}},
  \bibinfo{author}{\bibfnamefont{D.~a.} \bibnamefont{Romero}},
  \bibnamefont{and} \bibinfo{author}{\bibfnamefont{P.}~\bibnamefont{Horvath}},
  \bibinfo{journal}{Science} \textbf{\bibinfo{volume}{315}},
  \bibinfo{pages}{1709} (\bibinfo{year}{2007}), ISSN \bibinfo{issn}{0036-8075},
  \urlprefix\url{http://www.sciencemag.org/cgi/doi/10.1126/science.1138140}.

\bibitem[{\citenamefont{Makarova et~al.}(2011)\citenamefont{Makarova, Haft,
  Barrangou, Brouns, Charpentier, Horvath, Moineau, Mojica, Wolf, Yakunin
  et~al.}}]{Makarova2012}
\bibinfo{author}{\bibfnamefont{K.~S.} \bibnamefont{Makarova}},
  \bibinfo{author}{\bibfnamefont{D.~H.} \bibnamefont{Haft}},
  \bibinfo{author}{\bibfnamefont{R.}~\bibnamefont{Barrangou}},
  \bibinfo{author}{\bibfnamefont{S.~J.~J.} \bibnamefont{Brouns}},
  \bibinfo{author}{\bibfnamefont{E.}~\bibnamefont{Charpentier}},
  \bibinfo{author}{\bibfnamefont{P.}~\bibnamefont{Horvath}},
  \bibinfo{author}{\bibfnamefont{S.}~\bibnamefont{Moineau}},
  \bibinfo{author}{\bibfnamefont{F.~J.~M.} \bibnamefont{Mojica}},
  \bibinfo{author}{\bibfnamefont{Y.~I.} \bibnamefont{Wolf}},
  \bibinfo{author}{\bibfnamefont{A.~F.} \bibnamefont{Yakunin}},
  \bibnamefont{et~al.}, \bibinfo{journal}{Nature reviews. Microbiology}
  \textbf{\bibinfo{volume}{9}}, \bibinfo{pages}{467} (\bibinfo{year}{2011}),
  ISSN \bibinfo{issn}{1740-1534},
  \urlprefix\url{http://dx.doi.org/10.1038/nrmicro2577}.

\bibitem[{\citenamefont{Makarova et~al.}(2015)\citenamefont{Makarova, Wolf,
  Alkhnbashi, Costa, Shah, Saunders, Barrangou, Brouns, Charpentier, Haft
  et~al.}}]{Makarova2015}
\bibinfo{author}{\bibfnamefont{K.~S.} \bibnamefont{Makarova}},
  \bibinfo{author}{\bibfnamefont{Y.~I.} \bibnamefont{Wolf}},
  \bibinfo{author}{\bibfnamefont{O.~S.} \bibnamefont{Alkhnbashi}},
  \bibinfo{author}{\bibfnamefont{F.}~\bibnamefont{Costa}},
  \bibinfo{author}{\bibfnamefont{S.~A.} \bibnamefont{Shah}},
  \bibinfo{author}{\bibfnamefont{S.~J.} \bibnamefont{Saunders}},
  \bibinfo{author}{\bibfnamefont{R.}~\bibnamefont{Barrangou}},
  \bibinfo{author}{\bibfnamefont{S.~J.~J.} \bibnamefont{Brouns}},
  \bibinfo{author}{\bibfnamefont{E.}~\bibnamefont{Charpentier}},
  \bibinfo{author}{\bibfnamefont{D.~H.} \bibnamefont{Haft}},
  \bibnamefont{et~al.}, \bibinfo{journal}{Nature Reviews Microbiology}
  \textbf{\bibinfo{volume}{13}}, \bibinfo{pages}{722} (\bibinfo{year}{2015}),
  ISSN \bibinfo{issn}{1740-1526},
  \urlprefix\url{http://www.nature.com/doifinder/10.1038/nrmicro3569}.

\bibitem[{\citenamefont{Hargreaves et~al.}(2014)\citenamefont{Hargreaves,
  Flores, Lawley, and Clokie}}]{Hargreaves2014}
\bibinfo{author}{\bibfnamefont{K.~R.} \bibnamefont{Hargreaves}},
  \bibinfo{author}{\bibfnamefont{C.~O.} \bibnamefont{Flores}},
  \bibinfo{author}{\bibfnamefont{T.~D.} \bibnamefont{Lawley}},
  \bibnamefont{and} \bibinfo{author}{\bibfnamefont{M.~R.~J.}
  \bibnamefont{Clokie}}, \bibinfo{journal}{mBio} \textbf{\bibinfo{volume}{5}},
  \bibinfo{pages}{e01045} (\bibinfo{year}{2014}), ISSN
  \bibinfo{issn}{2150-7511},
  \urlprefix\url{http://mbio.asm.org/cgi/doi/10.1128/mBio.01045-13}.

\bibitem[{\citenamefont{McGhee and Sundin}(2012)}]{McGhee2012}
\bibinfo{author}{\bibfnamefont{G.~C.} \bibnamefont{McGhee}} \bibnamefont{and}
  \bibinfo{author}{\bibfnamefont{G.~W.} \bibnamefont{Sundin}},
  \bibinfo{journal}{PLoS ONE} \textbf{\bibinfo{volume}{7}}
  (\bibinfo{year}{2012}), ISSN \bibinfo{issn}{19326203}.

\bibitem[{\citenamefont{van Belkum et~al.}(2015)\citenamefont{van Belkum,
  Soriaga, LaFave, Akella, Veyrieras, Barbu, Shortridge, Blanc, Hannum,
  Zambardi et~al.}}]{VanBelkum2015}
\bibinfo{author}{\bibfnamefont{A.}~\bibnamefont{van Belkum}},
  \bibinfo{author}{\bibfnamefont{L.~B.} \bibnamefont{Soriaga}},
  \bibinfo{author}{\bibfnamefont{M.~C.} \bibnamefont{LaFave}},
  \bibinfo{author}{\bibfnamefont{S.}~\bibnamefont{Akella}},
  \bibinfo{author}{\bibfnamefont{J.-b.} \bibnamefont{Veyrieras}},
  \bibinfo{author}{\bibfnamefont{E.~M.} \bibnamefont{Barbu}},
  \bibinfo{author}{\bibfnamefont{D.}~\bibnamefont{Shortridge}},
  \bibinfo{author}{\bibfnamefont{B.}~\bibnamefont{Blanc}},
  \bibinfo{author}{\bibfnamefont{G.}~\bibnamefont{Hannum}},
  \bibinfo{author}{\bibfnamefont{G.}~\bibnamefont{Zambardi}},
  \bibnamefont{et~al.}, \bibinfo{journal}{mBio} \textbf{\bibinfo{volume}{6}},
  \bibinfo{pages}{1} (\bibinfo{year}{2015}), ISSN \bibinfo{issn}{2150-7511},
  \urlprefix\url{http://www.ncbi.nlm.nih.gov/pubmed/26604259}.

\bibitem[{\citenamefont{Makarova et~al.}(2013)\citenamefont{Makarova, Wolf, and
  Koonin}}]{Makarova2013}
\bibinfo{author}{\bibfnamefont{K.~S.} \bibnamefont{Makarova}},
  \bibinfo{author}{\bibfnamefont{Y.~I.} \bibnamefont{Wolf}}, \bibnamefont{and}
  \bibinfo{author}{\bibfnamefont{E.~V.} \bibnamefont{Koonin}},
  \bibinfo{journal}{Nucleic Acids Research} \textbf{\bibinfo{volume}{41}},
  \bibinfo{pages}{4360} (\bibinfo{year}{2013}), ISSN \bibinfo{issn}{03051048}.

\bibitem[{\citenamefont{Agari et~al.}(2010)\citenamefont{Agari, Sakamoto,
  Tamakoshi, Oshima, Kuramitsu, and Shinkai}}]{Agari2010}
\bibinfo{author}{\bibfnamefont{Y.}~\bibnamefont{Agari}},
  \bibinfo{author}{\bibfnamefont{K.}~\bibnamefont{Sakamoto}},
  \bibinfo{author}{\bibfnamefont{M.}~\bibnamefont{Tamakoshi}},
  \bibinfo{author}{\bibfnamefont{T.}~\bibnamefont{Oshima}},
  \bibinfo{author}{\bibfnamefont{S.}~\bibnamefont{Kuramitsu}},
  \bibnamefont{and} \bibinfo{author}{\bibfnamefont{A.}~\bibnamefont{Shinkai}},
  \bibinfo{journal}{Journal of Molecular Biology}
  \textbf{\bibinfo{volume}{395}}, \bibinfo{pages}{270} (\bibinfo{year}{2010}),
  ISSN \bibinfo{issn}{00222836},
  \urlprefix\url{http://dx.doi.org/10.1016/j.jmb.2009.10.057}.

\bibitem[{\citenamefont{Rath et~al.}(2015)\citenamefont{Rath, Amlinger, Rath,
  and Lundgren}}]{Rath2015}
\bibinfo{author}{\bibfnamefont{D.}~\bibnamefont{Rath}},
  \bibinfo{author}{\bibfnamefont{L.}~\bibnamefont{Amlinger}},
  \bibinfo{author}{\bibfnamefont{A.}~\bibnamefont{Rath}}, \bibnamefont{and}
  \bibinfo{author}{\bibfnamefont{M.}~\bibnamefont{Lundgren}},
  \bibinfo{journal}{Biochimie} \textbf{\bibinfo{volume}{117}},
  \bibinfo{pages}{119} (\bibinfo{year}{2015}), ISSN \bibinfo{issn}{61831638},
  \urlprefix\url{http://dx.doi.org/10.1016/j.biochi.2015.03.025}.

\bibitem[{\citenamefont{D{\'{i}}ez-Villase{\~{n}}or
  et~al.}(2010)\citenamefont{D{\'{i}}ez-Villase{\~{n}}or, Almendros,
  Garc{\'{i}}a-Mart{\'{i}}nez, and Mojica}}]{Diez-Villasenor2010}
\bibinfo{author}{\bibfnamefont{C.}~\bibnamefont{D{\'{i}}ez-Villase{\~{n}}or}},
  \bibinfo{author}{\bibfnamefont{C.}~\bibnamefont{Almendros}},
  \bibinfo{author}{\bibfnamefont{J.}~\bibnamefont{Garc{\'{i}}a-Mart{\'{i}}nez}},
  \bibnamefont{and} \bibinfo{author}{\bibfnamefont{F.~J.~M.}
  \bibnamefont{Mojica}}, \bibinfo{journal}{Microbiology}
  \textbf{\bibinfo{volume}{156}}, \bibinfo{pages}{1351} (\bibinfo{year}{2010}),
  ISSN \bibinfo{issn}{13500872}.

\bibitem[{\citenamefont{Horvath et~al.}(2008)\citenamefont{Horvath, Romero,
  Co{\^{u}}t{\'{e}}-Monvoisin, Richards, Deveau, Moineau, Boyaval, Fremaux, and
  Barrangou}}]{Horvath2008}
\bibinfo{author}{\bibfnamefont{P.}~\bibnamefont{Horvath}},
  \bibinfo{author}{\bibfnamefont{D.~A.} \bibnamefont{Romero}},
  \bibinfo{author}{\bibfnamefont{A.~C.}
  \bibnamefont{Co{\^{u}}t{\'{e}}-Monvoisin}},
  \bibinfo{author}{\bibfnamefont{M.}~\bibnamefont{Richards}},
  \bibinfo{author}{\bibfnamefont{H.}~\bibnamefont{Deveau}},
  \bibinfo{author}{\bibfnamefont{S.}~\bibnamefont{Moineau}},
  \bibinfo{author}{\bibfnamefont{P.}~\bibnamefont{Boyaval}},
  \bibinfo{author}{\bibfnamefont{C.}~\bibnamefont{Fremaux}}, \bibnamefont{and}
  \bibinfo{author}{\bibfnamefont{R.}~\bibnamefont{Barrangou}},
  \bibinfo{journal}{Journal of Bacteriology} \textbf{\bibinfo{volume}{190}},
  \bibinfo{pages}{1401} (\bibinfo{year}{2008}), ISSN \bibinfo{issn}{00219193}.

\bibitem[{\citenamefont{Grissa et~al.}(2007)\citenamefont{Grissa, Vergnaud, and
  Pourcel}}]{Grissa2007}
\bibinfo{author}{\bibfnamefont{I.}~\bibnamefont{Grissa}},
  \bibinfo{author}{\bibfnamefont{G.}~\bibnamefont{Vergnaud}}, \bibnamefont{and}
  \bibinfo{author}{\bibfnamefont{C.}~\bibnamefont{Pourcel}},
  \bibinfo{journal}{BMC bioinformatics} \textbf{\bibinfo{volume}{8}},
  \bibinfo{pages}{172} (\bibinfo{year}{2007}), ISSN \bibinfo{issn}{14712105}.

\bibitem[{\citenamefont{Hale et~al.}(2008)\citenamefont{Hale, Kleppe, Terns,
  and Terns}}]{Hale2008}
\bibinfo{author}{\bibfnamefont{C.}~\bibnamefont{Hale}},
  \bibinfo{author}{\bibfnamefont{K.}~\bibnamefont{Kleppe}},
  \bibinfo{author}{\bibfnamefont{R.~M.} \bibnamefont{Terns}}, \bibnamefont{and}
  \bibinfo{author}{\bibfnamefont{M.~P.} \bibnamefont{Terns}},
  \bibinfo{journal}{RNA (New York, N.Y.)} \textbf{\bibinfo{volume}{14}},
  \bibinfo{pages}{2572} (\bibinfo{year}{2008}), ISSN \bibinfo{issn}{1469-9001},
  \urlprefix\url{http://www.pubmedcentral.nih.gov/articlerender.fcgi?artid=2590957{\&}tool=pmcentrez{\&}rendertype=abstract}.

\bibitem[{\citenamefont{Levin et~al.}(2013)\citenamefont{Levin, Moineau,
  Bushman, and Barrangou}}]{Levin2013}
\bibinfo{author}{\bibfnamefont{B.~R.} \bibnamefont{Levin}},
  \bibinfo{author}{\bibfnamefont{S.}~\bibnamefont{Moineau}},
  \bibinfo{author}{\bibfnamefont{M.}~\bibnamefont{Bushman}}, \bibnamefont{and}
  \bibinfo{author}{\bibfnamefont{R.}~\bibnamefont{Barrangou}},
  \bibinfo{journal}{PLoS Genetics} \textbf{\bibinfo{volume}{9}}
  (\bibinfo{year}{2013}), ISSN \bibinfo{issn}{15537390}.

\bibitem[{\citenamefont{Marraffini and Sontheimer}(2008)}]{Marraffini2008}
\bibinfo{author}{\bibfnamefont{L.~A.} \bibnamefont{Marraffini}}
  \bibnamefont{and} \bibinfo{author}{\bibfnamefont{E.~J.}
  \bibnamefont{Sontheimer}}, \bibinfo{journal}{Science}
  \textbf{\bibinfo{volume}{322}}, \bibinfo{pages}{1843} (\bibinfo{year}{2008}),
  ISSN \bibinfo{issn}{0036-8075}, \eprint{NIHMS150003},
  \urlprefix\url{http://eutils.ncbi.nlm.nih.gov/entrez/eutils/elink.fcgi?dbfrom=pubmed{\&}id=19095942{\&}retmode=ref{\&}cmd=prlinks{\%}5Cnpapers2://publication/doi/10.1126/science.1165771}.

\bibitem[{\citenamefont{Bondy-Denomy and Davidson}(2014)}]{Bondy-Denomy2014}
\bibinfo{author}{\bibfnamefont{J.}~\bibnamefont{Bondy-Denomy}}
  \bibnamefont{and} \bibinfo{author}{\bibfnamefont{A.~R.}
  \bibnamefont{Davidson}}, \bibinfo{journal}{Trends in Microbiology}
  \textbf{\bibinfo{volume}{22}}, \bibinfo{pages}{218} (\bibinfo{year}{2014}),
  ISSN \bibinfo{issn}{18784380},
  \urlprefix\url{http://dx.doi.org/10.1016/j.tim.2014.01.007}.

\bibitem[{\citenamefont{Vale et~al.}(2015)\citenamefont{Vale, Lafforgue,
  Gatchitch, Gardan, Moineau, and Gandon}}]{Vale2015}
\bibinfo{author}{\bibfnamefont{P.~F.} \bibnamefont{Vale}},
  \bibinfo{author}{\bibfnamefont{G.}~\bibnamefont{Lafforgue}},
  \bibinfo{author}{\bibfnamefont{F.}~\bibnamefont{Gatchitch}},
  \bibinfo{author}{\bibfnamefont{R.}~\bibnamefont{Gardan}},
  \bibinfo{author}{\bibfnamefont{S.}~\bibnamefont{Moineau}}, \bibnamefont{and}
  \bibinfo{author}{\bibfnamefont{S.}~\bibnamefont{Gandon}},
  \bibinfo{journal}{Proceedings of the Royal Society B: Biological Sciences}
  \textbf{\bibinfo{volume}{282}}, \bibinfo{pages}{20151270}
  (\bibinfo{year}{2015}), ISSN \bibinfo{issn}{0962-8452},
  \urlprefix\url{http://dx.doi.org/10.1098/rspb.2015.1270{\%}5Cnhttp://rspb.royalsocietypublishing.org.
  http://rspb.royalsocietypublishing.org/lookup/doi/10.1098/rspb.2015.1270}.

\bibitem[{\citenamefont{Gophna et~al.}(2015)\citenamefont{Gophna, Kristensen,
  Wolf, Popa, Drevet, and Koonin}}]{Gophna2015}
\bibinfo{author}{\bibfnamefont{U.}~\bibnamefont{Gophna}},
  \bibinfo{author}{\bibfnamefont{D.~M.} \bibnamefont{Kristensen}},
  \bibinfo{author}{\bibfnamefont{Y.~I.} \bibnamefont{Wolf}},
  \bibinfo{author}{\bibfnamefont{O.}~\bibnamefont{Popa}},
  \bibinfo{author}{\bibfnamefont{C.}~\bibnamefont{Drevet}}, \bibnamefont{and}
  \bibinfo{author}{\bibfnamefont{E.~V.} \bibnamefont{Koonin}},
  \bibinfo{journal}{The ISME journal} \textbf{\bibinfo{volume}{9}},
  \bibinfo{pages}{2021} (\bibinfo{year}{2015}), ISSN \bibinfo{issn}{1751-7370},
  \urlprefix\url{http://dx.doi.org/10.1038/ismej.2015.20}.

\bibitem[{\citenamefont{Childs et~al.}(2012)\citenamefont{Childs, Held, Young,
  Whitaker, and Weitz}}]{Childs2012}
\bibinfo{author}{\bibfnamefont{L.~M.} \bibnamefont{Childs}},
  \bibinfo{author}{\bibfnamefont{N.~L.} \bibnamefont{Held}},
  \bibinfo{author}{\bibfnamefont{M.~J.} \bibnamefont{Young}},
  \bibinfo{author}{\bibfnamefont{R.~J.} \bibnamefont{Whitaker}},
  \bibnamefont{and} \bibinfo{author}{\bibfnamefont{J.~S.} \bibnamefont{Weitz}},
  \bibinfo{journal}{Evolution} \textbf{\bibinfo{volume}{66}},
  \bibinfo{pages}{2015} (\bibinfo{year}{2012}), ISSN \bibinfo{issn}{00143820},
  \urlprefix\url{http://doi.wiley.com/10.1111/j.1558-5646.2012.01595.x}.

\bibitem[{\citenamefont{Iranzo et~al.}(2013)\citenamefont{Iranzo, Lobkovsky,
  Wolf, and Koonin}}]{Iranzo2013}
\bibinfo{author}{\bibfnamefont{J.}~\bibnamefont{Iranzo}},
  \bibinfo{author}{\bibfnamefont{A.~E.} \bibnamefont{Lobkovsky}},
  \bibinfo{author}{\bibfnamefont{Y.~I.} \bibnamefont{Wolf}}, \bibnamefont{and}
  \bibinfo{author}{\bibfnamefont{E.~V.} \bibnamefont{Koonin}},
  \bibinfo{journal}{Journal of Bacteriology} \textbf{\bibinfo{volume}{195}},
  \bibinfo{pages}{3834} (\bibinfo{year}{2013}), ISSN \bibinfo{issn}{00219193}.

\bibitem[{\citenamefont{Bradde et~al.}(2017)\citenamefont{Bradde, Vucelja,
  Tesileanu, and Balasubramanian}}]{Bradde2017}
\bibinfo{author}{\bibfnamefont{S.}~\bibnamefont{Bradde}},
  \bibinfo{author}{\bibfnamefont{M.}~\bibnamefont{Vucelja}},
  \bibinfo{author}{\bibfnamefont{T.}~\bibnamefont{Tesileanu}},
  \bibnamefont{and}
  \bibinfo{author}{\bibfnamefont{V.}~\bibnamefont{Balasubramanian}},
  \bibinfo{journal}{PLOS Computational Biology} \textbf{\bibinfo{volume}{13}},
  \bibinfo{pages}{e1005486} (\bibinfo{year}{2017}), ISSN
  \bibinfo{issn}{1553-7358}, \eprint{1510.06082},
  \urlprefix\url{http://arxiv.org/abs/1510.06082
  http://dx.plos.org/10.1371/journal.pcbi.1005486}.

\bibitem[{\citenamefont{D{\'{i}}ez-Villase{\~{n}}or
  et~al.}(2013)\citenamefont{D{\'{i}}ez-Villase{\~{n}}or, Guzm{\'{a}}n,
  Almendros, Garc{\'{i}}a-Mart{\'{i}}nez, and Mojica}}]{Diez-Villasenor2013}
\bibinfo{author}{\bibfnamefont{C.}~\bibnamefont{D{\'{i}}ez-Villase{\~{n}}or}},
  \bibinfo{author}{\bibfnamefont{N.~M.} \bibnamefont{Guzm{\'{a}}n}},
  \bibinfo{author}{\bibfnamefont{C.}~\bibnamefont{Almendros}},
  \bibinfo{author}{\bibfnamefont{J.}~\bibnamefont{Garc{\'{i}}a-Mart{\'{i}}nez}},
  \bibnamefont{and} \bibinfo{author}{\bibfnamefont{F.~J.}
  \bibnamefont{Mojica}}, \bibinfo{journal}{RNA Biology}
  \textbf{\bibinfo{volume}{10}}, \bibinfo{pages}{792} (\bibinfo{year}{2013}),
  ISSN \bibinfo{issn}{1547-6286},
  \urlprefix\url{http://www.tandfonline.com/doi/full/10.4161/rna.24023}.

\bibitem[{\citenamefont{Jackson et~al.}(2017)\citenamefont{Jackson, McKenzie,
  Fagerlund, Kieper, Fineran, and Brouns}}]{Jackson2017}
\bibinfo{author}{\bibfnamefont{S.~A.} \bibnamefont{Jackson}},
  \bibinfo{author}{\bibfnamefont{R.~E.} \bibnamefont{McKenzie}},
  \bibinfo{author}{\bibfnamefont{R.~D.} \bibnamefont{Fagerlund}},
  \bibinfo{author}{\bibfnamefont{S.~N.} \bibnamefont{Kieper}},
  \bibinfo{author}{\bibfnamefont{P.~C.} \bibnamefont{Fineran}},
  \bibnamefont{and} \bibinfo{author}{\bibfnamefont{S.~J.~J.}
  \bibnamefont{Brouns}}, \bibinfo{journal}{Science}
  \textbf{\bibinfo{volume}{356}}, \bibinfo{pages}{eaal5056}
  (\bibinfo{year}{2017}), ISSN \bibinfo{issn}{0036-8075},
  \urlprefix\url{http://www.sciencemag.org/lookup/doi/10.1126/science.aal5056}.

\bibitem[{\citenamefont{Semenova et~al.}(2011)\citenamefont{Semenova, Jore,
  Datsenko, Semenova, Westra, Wanner, van~der Oost, Brouns, and
  Severinov}}]{Semenova2011}
\bibinfo{author}{\bibfnamefont{E.}~\bibnamefont{Semenova}},
  \bibinfo{author}{\bibfnamefont{M.~M.} \bibnamefont{Jore}},
  \bibinfo{author}{\bibfnamefont{K.~a.} \bibnamefont{Datsenko}},
  \bibinfo{author}{\bibfnamefont{A.}~\bibnamefont{Semenova}},
  \bibinfo{author}{\bibfnamefont{E.~R.} \bibnamefont{Westra}},
  \bibinfo{author}{\bibfnamefont{B.}~\bibnamefont{Wanner}},
  \bibinfo{author}{\bibfnamefont{J.}~\bibnamefont{van~der Oost}},
  \bibinfo{author}{\bibfnamefont{S.~J.~J.} \bibnamefont{Brouns}},
  \bibnamefont{and}
  \bibinfo{author}{\bibfnamefont{K.}~\bibnamefont{Severinov}},
  \bibinfo{journal}{Proceedings of the National Academy of Sciences of the
  United States of America} \textbf{\bibinfo{volume}{108}},
  \bibinfo{pages}{10098} (\bibinfo{year}{2011}), ISSN
  \bibinfo{issn}{0027-8424}.

\bibitem[{\citenamefont{Fischer et~al.}(2012)\citenamefont{Fischer, Maier,
  Stoll, Brendel, Fischer, Pfeiffer, Dyall-Smith, and
  Marchfelder}}]{Fischer2012}
\bibinfo{author}{\bibfnamefont{S.}~\bibnamefont{Fischer}},
  \bibinfo{author}{\bibfnamefont{L.~K.} \bibnamefont{Maier}},
  \bibinfo{author}{\bibfnamefont{B.}~\bibnamefont{Stoll}},
  \bibinfo{author}{\bibfnamefont{J.}~\bibnamefont{Brendel}},
  \bibinfo{author}{\bibfnamefont{E.}~\bibnamefont{Fischer}},
  \bibinfo{author}{\bibfnamefont{F.}~\bibnamefont{Pfeiffer}},
  \bibinfo{author}{\bibfnamefont{M.}~\bibnamefont{Dyall-Smith}},
  \bibnamefont{and}
  \bibinfo{author}{\bibfnamefont{A.}~\bibnamefont{Marchfelder}},
  \bibinfo{journal}{Journal of Biological Chemistry}
  \textbf{\bibinfo{volume}{287}}, \bibinfo{pages}{33351}
  (\bibinfo{year}{2012}), ISSN \bibinfo{issn}{00219258}.

\bibitem[{\citenamefont{Shah et~al.}(2013)\citenamefont{Shah, Erdmann, Mojica,
  and Garrett}}]{Shah2013}
\bibinfo{author}{\bibfnamefont{S.}~\bibnamefont{Shah}},
  \bibinfo{author}{\bibfnamefont{S.}~\bibnamefont{Erdmann}},
  \bibinfo{author}{\bibfnamefont{F.}~\bibnamefont{Mojica}}, \bibnamefont{and}
  \bibinfo{author}{\bibfnamefont{R.}~\bibnamefont{Garrett}},
  \bibinfo{journal}{RNA biology} \textbf{\bibinfo{volume}{10}},
  \bibinfo{pages}{891} (\bibinfo{year}{2013}), ISSN \bibinfo{issn}{1547-6286},
  \urlprefix\url{http://www.landesbioscience.com/journals/rnabiology/2012RNABIOL0169R.pdf}.

\bibitem[{\citenamefont{Pul et~al.}(2010)\citenamefont{Pul, Wurm, Arslan,
  Gei{\ss}en, Hofmann, and Wagner}}]{Pul2010}
\bibinfo{author}{\bibfnamefont{{\"{U}}.}~\bibnamefont{Pul}},
  \bibinfo{author}{\bibfnamefont{R.}~\bibnamefont{Wurm}},
  \bibinfo{author}{\bibfnamefont{Z.}~\bibnamefont{Arslan}},
  \bibinfo{author}{\bibfnamefont{R.}~\bibnamefont{Gei{\ss}en}},
  \bibinfo{author}{\bibfnamefont{N.}~\bibnamefont{Hofmann}}, \bibnamefont{and}
  \bibinfo{author}{\bibfnamefont{R.}~\bibnamefont{Wagner}},
  \bibinfo{journal}{Molecular Microbiology} \textbf{\bibinfo{volume}{75}},
  \bibinfo{pages}{1495} (\bibinfo{year}{2010}), ISSN \bibinfo{issn}{13652958}.

\bibitem[{\citenamefont{Zoephel and Randau}(2013)}]{Zoephel2013}
\bibinfo{author}{\bibfnamefont{J.}~\bibnamefont{Zoephel}} \bibnamefont{and}
  \bibinfo{author}{\bibfnamefont{L.}~\bibnamefont{Randau}},
  \bibinfo{journal}{Biochemical Society Transactions}
  \textbf{\bibinfo{volume}{41}}, \bibinfo{pages}{1459} (\bibinfo{year}{2013}),
  ISSN \bibinfo{issn}{0300-5127},
  \urlprefix\url{http://biochemsoctrans.org/lookup/doi/10.1042/BST20130129}.

\bibitem[{\citenamefont{Strotskaya et~al.}(2017)\citenamefont{Strotskaya,
  Savitskaya, Metlitskaya, Morozova, Datsenko, Semenova, and
  Severinov}}]{Strotskaya2017}
\bibinfo{author}{\bibfnamefont{A.}~\bibnamefont{Strotskaya}},
  \bibinfo{author}{\bibfnamefont{E.}~\bibnamefont{Savitskaya}},
  \bibinfo{author}{\bibfnamefont{A.}~\bibnamefont{Metlitskaya}},
  \bibinfo{author}{\bibfnamefont{N.}~\bibnamefont{Morozova}},
  \bibinfo{author}{\bibfnamefont{K.~A.} \bibnamefont{Datsenko}},
  \bibinfo{author}{\bibfnamefont{E.}~\bibnamefont{Semenova}}, \bibnamefont{and}
  \bibinfo{author}{\bibfnamefont{K.}~\bibnamefont{Severinov}},
  \bibinfo{journal}{Nucleic Acids Research} p. \bibinfo{pages}{gkx042}
  (\bibinfo{year}{2017}), ISSN \bibinfo{issn}{0305-1048},
  \urlprefix\url{http://nar.oxfordjournals.org/lookup/doi/10.1093/nar/gkx042}.

\bibitem[{\citenamefont{Berezovskaya et~al.}(2014)\citenamefont{Berezovskaya,
  Wolf, Koonin, and Karev}}]{Berezovskaya2014}
\bibinfo{author}{\bibfnamefont{F.~S.} \bibnamefont{Berezovskaya}},
  \bibinfo{author}{\bibfnamefont{Y.~I.} \bibnamefont{Wolf}},
  \bibinfo{author}{\bibfnamefont{E.~V.} \bibnamefont{Koonin}},
  \bibnamefont{and} \bibinfo{author}{\bibfnamefont{G.~P.} \bibnamefont{Karev}},
  \bibinfo{journal}{Biology direct} \textbf{\bibinfo{volume}{9}},
  \bibinfo{pages}{13} (\bibinfo{year}{2014}), ISSN \bibinfo{issn}{1745-6150},
  \urlprefix\url{http://biologydirect.biomedcentral.com/articles/10.1186/1745-6150-9-13
  http://www.ncbi.nlm.nih.gov/pubmed/24986220{\%}5Cnhttp://www.biologydirect.com/content/pdf/1745-6150-9-13.pdf
  http://www.ncbi.nlm.nih.gov/pubmed/24986220
  http://www.pubmedcentral.nih.gov/}.

\bibitem[{\citenamefont{Rao et~al.}(2017)\citenamefont{Rao, Chin, and
  Ensminger}}]{Rao2017}
\bibinfo{author}{\bibfnamefont{C.}~\bibnamefont{Rao}},
  \bibinfo{author}{\bibfnamefont{D.}~\bibnamefont{Chin}}, \bibnamefont{and}
  \bibinfo{author}{\bibfnamefont{A.~W.} \bibnamefont{Ensminger}},
  \bibinfo{journal}{bioRxiv}  (\bibinfo{year}{2017}).

\bibitem[{\citenamefont{Semenova et~al.}(2016)\citenamefont{Semenova,
  Savitskaya, Musharova, Strotskaya, Vorontsova, Datsenko, Logacheva, and
  Severinov}}]{Semenova2016}
\bibinfo{author}{\bibfnamefont{E.}~\bibnamefont{Semenova}},
  \bibinfo{author}{\bibfnamefont{E.}~\bibnamefont{Savitskaya}},
  \bibinfo{author}{\bibfnamefont{O.}~\bibnamefont{Musharova}},
  \bibinfo{author}{\bibfnamefont{A.}~\bibnamefont{Strotskaya}},
  \bibinfo{author}{\bibfnamefont{D.}~\bibnamefont{Vorontsova}},
  \bibinfo{author}{\bibfnamefont{K.~A.} \bibnamefont{Datsenko}},
  \bibinfo{author}{\bibfnamefont{M.~D.} \bibnamefont{Logacheva}},
  \bibnamefont{and}
  \bibinfo{author}{\bibfnamefont{K.}~\bibnamefont{Severinov}},
  \bibinfo{journal}{Proceedings of the National Academy of Sciences of the
  United States of America} \textbf{\bibinfo{volume}{113}},
  \bibinfo{pages}{7626} (\bibinfo{year}{2016}), ISSN \bibinfo{issn}{1091-6490
  (Electronic)}.

\bibitem[{\citenamefont{Weinberger et~al.}(2012)\citenamefont{Weinberger, Wolf,
  Lobkovsky, Gilmore, and Koonin}}]{Weinberger2012}
\bibinfo{author}{\bibfnamefont{A.~D.} \bibnamefont{Weinberger}},
  \bibinfo{author}{\bibfnamefont{Y.~I.} \bibnamefont{Wolf}},
  \bibinfo{author}{\bibfnamefont{A.~E.} \bibnamefont{Lobkovsky}},
  \bibinfo{author}{\bibfnamefont{M.~S.} \bibnamefont{Gilmore}},
  \bibnamefont{and} \bibinfo{author}{\bibfnamefont{E.~V.}
  \bibnamefont{Koonin}}, \bibinfo{journal}{mBio} \textbf{\bibinfo{volume}{3}},
  \bibinfo{pages}{1} (\bibinfo{year}{2012}), ISSN \bibinfo{issn}{21507511}.

\bibitem[{\citenamefont{Kunin et~al.}(2007)\citenamefont{Kunin, Sorek, and
  Hugenholtz}}]{Kunin2007}
\bibinfo{author}{\bibfnamefont{V.}~\bibnamefont{Kunin}},
  \bibinfo{author}{\bibfnamefont{R.}~\bibnamefont{Sorek}}, \bibnamefont{and}
  \bibinfo{author}{\bibfnamefont{P.}~\bibnamefont{Hugenholtz}},
  \bibinfo{journal}{Genome biology} \textbf{\bibinfo{volume}{8}},
  \bibinfo{pages}{R61} (\bibinfo{year}{2007}), ISSN \bibinfo{issn}{1474-760X},
  \urlprefix\url{http://genomebiology.biomedcentral.com/articles/10.1186/gb-2007-8-4-r61}.

\bibitem[{\citenamefont{Brouns et~al.}(1993)\citenamefont{Brouns, Jore,
  Lundgren, Westra, Slijkhuis, Snijders, Dickman, Makarova, Koonin, and {Van
  Der Oost}}}]{Brouns1993}
\bibinfo{author}{\bibfnamefont{S.~J.~J.} \bibnamefont{Brouns}},
  \bibinfo{author}{\bibfnamefont{M.~M.} \bibnamefont{Jore}},
  \bibinfo{author}{\bibfnamefont{M.}~\bibnamefont{Lundgren}},
  \bibinfo{author}{\bibfnamefont{E.~R.} \bibnamefont{Westra}},
  \bibinfo{author}{\bibfnamefont{R.~J.~H.} \bibnamefont{Slijkhuis}},
  \bibinfo{author}{\bibfnamefont{A.~P.~L.} \bibnamefont{Snijders}},
  \bibinfo{author}{\bibfnamefont{M.~J.} \bibnamefont{Dickman}},
  \bibinfo{author}{\bibfnamefont{K.~S.} \bibnamefont{Makarova}},
  \bibinfo{author}{\bibfnamefont{E.~V.} \bibnamefont{Koonin}},
  \bibnamefont{and} \bibinfo{author}{\bibfnamefont{J.}~\bibnamefont{{Van Der
  Oost}}}, \bibinfo{journal}{Cancer Epidemiology Biomarkers and Prevention}
  \textbf{\bibinfo{volume}{2}}, \bibinfo{pages}{531} (\bibinfo{year}{1993}),
  ISSN \bibinfo{issn}{10559965}, \eprint{20}.

\bibitem[{\citenamefont{Wilson and Hippel}(1995)}]{Wilson1995}
\bibinfo{author}{\bibfnamefont{K.~S.} \bibnamefont{Wilson}} \bibnamefont{and}
  \bibinfo{author}{\bibfnamefont{P.~H.~V.} \bibnamefont{Hippel}},
  \bibinfo{journal}{Biochemistry} \textbf{\bibinfo{volume}{92}},
  \bibinfo{pages}{8793} (\bibinfo{year}{1995}), ISSN \bibinfo{issn}{0027-8424}.

\bibitem[{\citenamefont{Farnham and Platt}(1981)}]{Farnham1981}
\bibinfo{author}{\bibfnamefont{P.~J.} \bibnamefont{Farnham}} \bibnamefont{and}
  \bibinfo{author}{\bibfnamefont{T.}~\bibnamefont{Platt}},
  \bibinfo{journal}{Nucleic Acids Research} \textbf{\bibinfo{volume}{9}},
  \bibinfo{pages}{563} (\bibinfo{year}{1981}), ISSN \bibinfo{issn}{03051048}.

\bibitem[{\citenamefont{Han and Deem}(2017)}]{Han2017}
\bibinfo{author}{\bibfnamefont{P.}~\bibnamefont{Han}} \bibnamefont{and}
  \bibinfo{author}{\bibfnamefont{M.~W.} \bibnamefont{Deem}},
  \bibinfo{journal}{Journal of The Royal Society Interface}
  \textbf{\bibinfo{volume}{14}}, \bibinfo{pages}{20160905}
  (\bibinfo{year}{2017}), ISSN \bibinfo{issn}{1742-5689},
  \urlprefix\url{http://rsif.royalsocietypublishing.org/lookup/doi/10.1098/rsif.2016.0905}.

\bibitem[{\citenamefont{Kuo and Ochman}(2010)}]{Kuo2009}
\bibinfo{author}{\bibfnamefont{C.~H.} \bibnamefont{Kuo}} \bibnamefont{and}
  \bibinfo{author}{\bibfnamefont{H.}~\bibnamefont{Ochman}},
  \bibinfo{journal}{Genome Biology and Evolution} \textbf{\bibinfo{volume}{1}},
  \bibinfo{pages}{145} (\bibinfo{year}{2010}), ISSN \bibinfo{issn}{1759-6653},
  \urlprefix\url{https://academic.oup.com/gbe/article-lookup/doi/10.1093/gbe/evp016}.

\bibitem[{\citenamefont{Vercoe et~al.}(2013)\citenamefont{Vercoe, Chang, Dy,
  Taylor, Gristwood, Clulow, Richter, Przybilski, Pitman, and
  Fineran}}]{Vercoe2013}
\bibinfo{author}{\bibfnamefont{R.~B.} \bibnamefont{Vercoe}},
  \bibinfo{author}{\bibfnamefont{J.~T.} \bibnamefont{Chang}},
  \bibinfo{author}{\bibfnamefont{R.~L.} \bibnamefont{Dy}},
  \bibinfo{author}{\bibfnamefont{C.}~\bibnamefont{Taylor}},
  \bibinfo{author}{\bibfnamefont{T.}~\bibnamefont{Gristwood}},
  \bibinfo{author}{\bibfnamefont{J.~S.} \bibnamefont{Clulow}},
  \bibinfo{author}{\bibfnamefont{C.}~\bibnamefont{Richter}},
  \bibinfo{author}{\bibfnamefont{R.}~\bibnamefont{Przybilski}},
  \bibinfo{author}{\bibfnamefont{A.~R.} \bibnamefont{Pitman}},
  \bibnamefont{and} \bibinfo{author}{\bibfnamefont{P.~C.}
  \bibnamefont{Fineran}}, \bibinfo{journal}{PLoS Genetics}
  \textbf{\bibinfo{volume}{9}} (\bibinfo{year}{2013}), ISSN
  \bibinfo{issn}{15537390}.

\bibitem[{\citenamefont{Wei et~al.}(2015)\citenamefont{Wei, Terns, and
  Terns}}]{Wei2015}
\bibinfo{author}{\bibfnamefont{Y.}~\bibnamefont{Wei}},
  \bibinfo{author}{\bibfnamefont{R.~M.} \bibnamefont{Terns}}, \bibnamefont{and}
  \bibinfo{author}{\bibfnamefont{M.~P.} \bibnamefont{Terns}},
  \bibinfo{journal}{Genes {\&} Development} \textbf{\bibinfo{volume}{29}},
  \bibinfo{pages}{356} (\bibinfo{year}{2015}), ISSN \bibinfo{issn}{0890-9369},
  \urlprefix\url{http://genesdev.cshlp.org/lookup/doi/10.1101/gad.257550.114}.

\bibitem[{\citenamefont{Levy et~al.}(2015)\citenamefont{Levy, Goren, Yosef,
  Auster, Manor, Amitai, Edgar, Qimron, and Sorek}}]{Levy2015}
\bibinfo{author}{\bibfnamefont{A.}~\bibnamefont{Levy}},
  \bibinfo{author}{\bibfnamefont{M.~G.} \bibnamefont{Goren}},
  \bibinfo{author}{\bibfnamefont{I.}~\bibnamefont{Yosef}},
  \bibinfo{author}{\bibfnamefont{O.}~\bibnamefont{Auster}},
  \bibinfo{author}{\bibfnamefont{M.}~\bibnamefont{Manor}},
  \bibinfo{author}{\bibfnamefont{G.}~\bibnamefont{Amitai}},
  \bibinfo{author}{\bibfnamefont{R.}~\bibnamefont{Edgar}},
  \bibinfo{author}{\bibfnamefont{U.}~\bibnamefont{Qimron}}, \bibnamefont{and}
  \bibinfo{author}{\bibfnamefont{R.}~\bibnamefont{Sorek}},
  \bibinfo{journal}{Nature} \textbf{\bibinfo{volume}{520}},
  \bibinfo{pages}{505} (\bibinfo{year}{2015}), ISSN \bibinfo{issn}{0028-0836},
  \urlprefix\url{http://www.nature.com/doifinder/10.1038/nature14302}.

\bibitem[{\citenamefont{Yosef et~al.}(2012)\citenamefont{Yosef, Goren, and
  Qimron}}]{Yosef2012}
\bibinfo{author}{\bibfnamefont{I.}~\bibnamefont{Yosef}},
  \bibinfo{author}{\bibfnamefont{M.~G.} \bibnamefont{Goren}}, \bibnamefont{and}
  \bibinfo{author}{\bibfnamefont{U.}~\bibnamefont{Qimron}},
  \bibinfo{journal}{Nucleic Acids Research} \textbf{\bibinfo{volume}{40}},
  \bibinfo{pages}{5569} (\bibinfo{year}{2012}), ISSN \bibinfo{issn}{03051048}.

\bibitem[{\citenamefont{Jiang et~al.}(2013)\citenamefont{Jiang, Maniv, Arain,
  Wang, Levin, and Marraffini}}]{Jiang2013}
\bibinfo{author}{\bibfnamefont{W.}~\bibnamefont{Jiang}},
  \bibinfo{author}{\bibfnamefont{I.}~\bibnamefont{Maniv}},
  \bibinfo{author}{\bibfnamefont{F.}~\bibnamefont{Arain}},
  \bibinfo{author}{\bibfnamefont{Y.}~\bibnamefont{Wang}},
  \bibinfo{author}{\bibfnamefont{B.~R.} \bibnamefont{Levin}}, \bibnamefont{and}
  \bibinfo{author}{\bibfnamefont{L.~A.} \bibnamefont{Marraffini}},
  \bibinfo{journal}{PLoS Genetics} \textbf{\bibinfo{volume}{9}},
  \bibinfo{pages}{e1003844} (\bibinfo{year}{2013}), ISSN
  \bibinfo{issn}{1553-7404},
  \urlprefix\url{http://dx.plos.org/10.1371/journal.pgen.1003844}.

\end{thebibliography}

\end{document}